\documentstyle[epsfig,subfigure]{mn}
 at 10truept
\title
     [Probability Distribution Function in Redshift Space]
{\vglue-3.0truecm
\centerline{\it For submission to Monthly Notices}
\vglue 2.5truecm
      Evolution of the probability distribution function of galaxies
      in redshift-space
\author
     [P.I.R Watts \& A.N. Taylor]
     {P.I.R. Watts \& A.N. Taylor\\
     Institute for Astronomy,
     University of Edinburgh,
     Royal Observatory,
     Blackford Hill,
     Edinburgh,
     U.K.\\
     pirw@roe.ac.uk, ant@roe.ac.uk}}
\def\bib{\parskip=0pt\par\noindent\hangindent\parindent
    \parskip =2ex plus .5ex minus .1ex}

\newcommand{\be}{\begin{equation}}
\newcommand{\ee}{\end{equation}}
\newcommand{\ba}{\begin{eqnarray}}
\newcommand{\ea}{\end{eqnarray}}

\newcommand{\pd}{\partial}
\newcommand{\fg}{{\mathcal G}}
\newcommand{\nn}{\nonumber \\}
\newcommand{\nnb}{\begin{displaymath}}
\newcommand{\nne}{\end{displaymath}}
\newcommand{\de}{\partial}
\newcommand{\x}{\mbox{\boldmath $x$}}
\newcommand{\y}{\mbox{\boldmath $y$}}
\newcommand{\s}{\mbox{\boldmath $s$}}
\newcommand{\g}{\mbox{\boldmath $g$}}
\newcommand{\E}{\mbox{\boldmath $E$}}
\newcommand{\C}{\mbox{\boldmath $C$}}

\newcommand{\vb}{\mbox{\boldmath $v$}}

\newcommand{\r}{\mbox{\boldmath $r$}}

\newcommand{\nablab}{\mbox{\boldmath $\nabla$}}
\newcommand{\sk}{$S^{s}_{3}$ }

\newcommand{\lll}{\left(}
\newcommand{\rrr}{\right)}

\newcommand{\etab}{\mbox{\boldmath $\eta$}}
\newcommand{\rgl}{\rangle}
\newcommand{\lgl}{\langle}

\newcommand{\bog}[1]{
    \mbox{\boldmath $ #1 $}
}

\twocolumn

\begin{document}

\maketitle

\begin{abstract}
We present a new analytic calculation for the redshift-space
evolution of the 1-point galaxy Probability Distribution Function
(PDF). The nonlinear evolution of the matter density field is treated
by second-order Eulerian perturbation theory and transformed to the
galaxy density field via a second-order local biasing scheme. We then
transform the galaxy density field to redshift space, again to second
order. Our method uses an exact statistical treatment based on the
Chapman--Kolmogorov equation to propagate the probability distribution
of the initial mass field to the final redshifted galaxy density
field. We derive the moment generating function of the PDF and use it
to find a new, closed-form expression for the skewness of the
redshifted galaxy distribution. We show that our formalism is general
enough to allow a non-deterministic (or stochastic) biasing
prescription. We demonstrate the dependence of the redshift space PDF
on cosmological and biasing parameters. Our results are compared with
existing models for the PDF in redshift space and with the results of
biased N-body simulations. We find that our PDF accurately models the
redshift space evolution and the nonlinear biasing.
\end{abstract}


\section{Introduction}

One of the central goals of modern cosmology is to gain a
quantitative understanding of the nature of the observed galaxy
distribution.  In order that such an understanding be complete it
must incorporate an explanation of both the evolution of
large-scale structure in the universe and of the relationship
between galaxies and the underlying dark mass.

In the standard paradigm the origin of structure can be traced
back to small perturbations in the mass density imprinted during
an inflationary phase. These perturbations subsequently grow in
amplitude through gravitational instability. This prescription is
well tested in N-body simulations and can be used to make useful
predictions about the clustering history of dark matter. Many of
the details, however, are not well understood.

The picture becomes more complex when one considers the galaxy
distribution. Firstly, the astrophysical processes that govern the
formation of the galaxies themselves introduce a nontrivial
relationship between the galaxies and the dark matter. This
relationship is generally termed galaxy bias. The biasing relation may
prove to be simple to describe, depending only upon the local density
in dark matter at a given point (eg Coles 1993). Conversely it could
be extremely complex, depending on the larger scale environment, tidal
fields, feedback and entire history of the galaxy formation site. This
could lead to non-local (Bower et al. 1993) or non-deterministic forms
of biasing (Pen 1998, Dekel \& Lahav 1999). Secondly, the observed
galaxy distribution is distorted in 3D redshift catalogues due to the
line of sight components of the galaxy peculiar motions adding to the
general Hubble expansion (Kaiser 1987). These redshift distortions, if
not properly modeled, may lead to misinterpretation when comparing
theory with observation.

A complete specification of galaxy clustering may only truly be given
by the full set of galaxy N-point correlation functions. Such
functions form the solution to the so-called BBGKY system of dynamical
equations.  This approach, pioneered in the 1970s by Peebles and co
workers (Davis \& Peebles 1977, Groth \& Peebles 1977, Fry \& Peebles
1978, Peebles 1980), has met with little success in practice. From a
theoretical perspective, a closed solution to the BBGKY hierarchy has
never been found and remains a thorny problem. Observationally,
measurements of the correlation functions have been restricted to the
lowest few orders.

An alternative description, and the focus of this work, may be given
by the Probability Distribution Function (PDF) of a random
field. Strictly, we investigate the PDF at a single spatial location
(the 1-point PDF), but in general the PDFs are a family of N-point
distribution functions for the probability at N positions in space.
The random fields of interest are the density fields of both dark
matter and galaxies.

PDFs are useful in cosmology because, in principle, they encode much
of the information contained within the full hierarchy of correlation
functions, thus providing valuable information about gravitational
evolution and initial conditions. Furthermore the discrete analogy of
the 1-point PDF, the counts in cells statistic, is a relatively
straightforward quantity to measure from galaxy surveys (Hamilton
1985, Alimi et al. 1990, Szapudi et al. 1992, 1996, Gazta\~{n}aga
1992, 1994, Bouchet et al. 1993, Kim \& Strauss 1998), allowing for
easy comparison between theory and observation.

The shape of the PDF is strongly influenced by the non-linear
effects of gravitational instability. A cosmic field that is
initially Gaussian random, the generic prediction of most
inflationary models, will remain so only when its evolution is
linear. When the evolution becomes nonlinear, the higher-order
moments of the field (more correctly the cumulants) become
non-zero for the first time resulting in a PDF that may be
strongly skewed about its mean and cuspy around its peak.

Attempts to describe theoretically this evolution have been
numerous (Fry 1985, Coles \& Jones 1991, Bernardeau 1992, 1994,
1996, Bernardeau \& Kofman 1995, Juszkiewicz et al. 1995, Colombi
et al. 1997, Gaztanaga et al. 1999, Taylor \& Watts 2000) with the
result that, at least for quasi-linear evolution of the matter
field, the behavior of the PDF is quite well understood. Since the
density field of galaxies inferred from a galaxy redshift survey
may not be a good representation of the underlying dark matter it
is unlikely that the PDF of the dark matter density field is a
realistic approximation for that of galaxies. Just as nonlinear
gravitational evolution drives the PDF away from Gaussianity, so
too will nonlinear bias and redshift distortions. The challenge is
to separate out each of the effects and quantify the evolution of
the PDF in terms of a few essential parameters.

There have been significantly fewer attempts to calculate the
evolution of the PDF in redshift space than in real space. Most
notably, Hui et al. (2000) extended the work of Bernardeau \&
Kofman (1995) to calculate the evolution using the Zel'dovich
approximation. Despite being an excellent
approximation for the nonlinear gravitational dynamics
Zel'dovich's solution does not produce a good fit to the PDF of
N-body simulations. This is principally due to the formation of
caustics in the density field at shell crossing. The result for
the Zel'dovich PDF is an asymptotic high density tail that falls
off like $\delta^{-3}$ in both real and redshift space. Tails like
these are not observed in simulations.

Another way of incorporating the effect of redshift distortions
and bias is to use the Edgeworth expansion (Juszkiewicz et al. 1995,
Bernardeau \& Kofman 1995). In this approximation, the PDF 
is reconstructed from
the moments of the field which must be calculated separately using
perturbation theory. Such a calculation was done by Juszkiewicz et
al. (1993; but see Hivon et al. 1995 for more details), who used
second order Lagrangian perturbation theory (Bouchet et al. 1995) to map the
skewness of an unbiased field into redshift space. They did not,
however, try to use the skewness to evolve the PDF via the Edgeworth expansion.

A problem with using the Edgeworth expansion is its tendency to
produce unphysical features in the PDF. In particular is the
appearance of negative probabilities at low densities and the
formation of unphysical ``wiggles'' in the high density tail of
the distribution when the variance grows. These problems can be
overcome by reconstructing the PDF using a Gamma expansion
(Gazta\~{n}aga et al. 1999) which shows a markedly better agreement
with N-body data in real space. 

In this paper we extend the formalism developed in a earlier work
(Taylor \& Watts 2000, henceforth TW2000) where we showed how the
PDF transforms when the matter density field is propagated to
second-order. This method is based on an exact propagation of
Gaussian initial probabilities using the Chapman--Kolmogorov
equation from statistical physics (e.g van Kampen 1992). In this
paper we develop this approach to include the transformation to a
local, second-order biased galaxy distribution, and apply
second-order Eulerian perturbation theory to describe the mapping
into redshift-space. We show how the method can be naturally
extended to incorporate a stochastic (or hidden variable) bias,
although we leave detailed calculations for a later paper (Watts
\& Taylor in preparation). Our calculation also provides a new
analytic solution for the skewness in redshift-space. This result
is different to that of Hivon et al. (1995) who estimated the
skewness by inverse transform of the bispectrum. However they
where unable to find a closed-form solution, and did not include
the effects of nonlinear galaxy biasing. Here we derive a
closed-form solution for the skewness, including second-order
bias, and find that the nonlinear bias has a large effect.

The layout of this paper is as follows. In \S 2 we discuss the
derivation of the 1-point Probability Distribution Function. The new
result for the skewness, and a discussion of stochastic biasing may
also be found here. In \S3 we illustrate the various dependencies on
cosmological parameters of the shape of the PDF. Our results are
compared with N-body simulations and other approximations in \S4, 
conclusions are given in \S 5. 

\section{The galaxy distribution function in redshift-space}

\subsection{Redshift distortions and biasing in second-order perturbation
theory.}

\subsubsection{Second-order perturbation theory}

In Eulerian perturbation theory the density field $\delta(\x,t)$
at real space comoving position $\x$ and time $t$ in a flat
Universe can be expanded into a series of separable functions;
\be
    \delta(\x,t)=\sum_{n=1}^\infty \delta_n(\x,t)
            =\sum_{n=1}^\infty D_n(t) \varepsilon_n(\x),
\ee 
In curved space this expansion is only separable to second-order. 
The evolution of $\delta(\x,t)$ can therefore be traced into
the nonlinear regime by solving the fluid and Poisson's equations for
each order in the perturbation expansion (Peebles 1980, Juszkiewicz
1981, Vishniac 1983, Fry 1984, Bouchet et al. 1992, Bouchet et al.
1995);
\ba
    \delta&=&\delta_1 + \delta_2 \nn
    &=& \delta_{1}+\frac{1}{3}(2-\kappa)\delta_1^2
 -\etab . \g+\frac{1}{2}(1+\kappa)E^{2},
\label{nl}
\ea
where
\be
    \etab(\x,t) = \nablab \delta_1(\x,t),
\ee
is the gradient of the linear density field and
\be
    \g(\x,t) = - \nablab \nabla^{-2} \delta_1(\x,t),
\ee
is the linear peculiar gravity field\footnote{In this paper we define
 $4 \pi G \rho_0=3/2\Omega H^{2} = 1$ and the expansion parameter $a(t)=1$,
since our final distribution function will be dimensionless.}
where $\nabla^{-2}$ is the inverse Laplacian. The trace-free tidal
tensor is given by
\be
    E_{ij}(\x,t) =( \nabla_i \nabla_j \nabla^{-2}  -
    \frac{1}{3}  \delta_{ij} )\, \delta_1(\x,t).
\ee

\subsubsection{Redshift space distortions}
To determine the redshift space density field one must consider the
mapping from the real space comoving position to redshift space comoving
position, $\s$, given by (see the excellent review by Hamilton, 1998,
for a full account)
\be
\s = \r + \hat{\r} u \ee where $u=\hat{\r}\cdot\vb(\r,t)/H_{0}$ is
the projection of the velocity field along the line of sight. In
the distant observer approximation (Kaiser 1987) the result can be written as a
series in $u$ and $\delta$,
\be
\delta^{s}(\s,t)=\sum_{n=0}^\infty
\frac{(-u)^{n}}{n!}\frac{\de^n}{\de s^n}
    \left(\frac{\delta(\s,t)-u'}{1+u'}\right),
\label{series}
\ee
where $'\equiv \partial/\partial r$ and all quantities are evaluated
in redshift-space coordinates.

Expanding this to second order we find
\be
	\delta^{s} = \delta - u' - [u(\delta-u')]'.
\ee

\subsubsection{Galaxy bias}
Next we assume that the density field of galaxies, smoothed on some
scale $R$, is some local function of the
underlying smoothed field of dark mass. Following Fry \& Gazta\~{n}aga
(1993) we
expand the galaxy density in powers of $\delta$;
\be \delta_{g}
= \sum_{n} \frac{b_{n}}{n!}\delta^{n}.
\label{bias}
\ee The coefficients in this expansion, $b_{n}$, are the bias
parameters.  We make no assumptions about the biasing function other
than that it is local and that it may be expanded in a Taylor
series. This is a deterministic Eulerian biasing scheme, but can be
generalised to a stochastic Eulerian biasing scheme to allow for the
hidden effects of galaxy formation (Pen 1998, Dekel \& Lahav 1999).
In Section \ref{SEB} we show how our intrinsically probabilistic
approach can be used to incorporate a stochastic Eularian bias. Other
alternatives for bias are nonlocal Eulerian biases (Bower et al. 1993)
and Lagrangian biasing schemes, which are intrinsically nonlocal in
Eulerian coordinates.  The latter is possibly the most natural to
arise from following halos in the Press-Schechter approach to galaxy
formation (Press \& Schechter 1974).  Our approach is more
phenomenological and we do not consider these possibilities here.

Combining equations (\ref{series}) and (\ref{bias}) gives to second order
(Heavens et al. 1998)
\be
\delta^{s}_{g}  =  b_1 \delta - u' + \frac{b_{2}}{2}(\delta^{2}-\sigma_{0}^2)
  +  u'^{2} - b_{1}\delta u'-b_{1}\delta' u + u u''
\label{heavens}
\ee
where all quantities are evaluated at $\s$.

In order that this expression has the correct expectation value,
\mbox{$\langle \delta_g^s \rangle = 0$},  we have set
\mbox{$b_0 = -b_2 \sigma_0^2/2$}.
Finally we rewrite equation (\ref{heavens}) in terms of the
linear density, gravity and tidal fields we shall use in later
calculations;
\be
\delta^{s}_{g} = \delta^{s}_{g,1} + \delta^{s}_{g,2} + \Delta^{s}_{g,1} +
    \Delta^{s}_{g,2},
\label{final}
\ee
where
\be
\delta^{s}_{g,1} = b_{1}\delta_{1},
\label{del1}
\ee
\be
    \delta^{s}_{g,2} =
    b_1 \delta_2  + \frac{b_{2}}{2}(\delta_{1}^{2} - \sigma_{0}^{2}),
\ee
\be
\Delta^{s}_{g,1} = f_1 \left(E_{zz} + \frac{\delta_{1}}{3}\right),
\ee
\begin{eqnarray}
\Delta^{s}_{g,2} & = &  \Delta^{s 2}_{g,1} + \delta^{s}_{g,1}\Delta^{s}_{g,1}
     -  \frac{f_{2}}{3}\kappa\lll \Pi_{zz}^{\delta} +
        \frac{\delta_{1}^{2}}{3}\rrr  \nn
    &+& \frac{f_{2}}{2}\kappa\lll\Pi_{zz}^{E}+ \frac{E^{2}}{3}\rrr
     -  f_{1} g_{z}\lll b_{1}\eta_{z} + f_{1} F_{zzz}\rrr.
\label{Del2}
\end{eqnarray}
The first two terms in equation (\ref{final}) deal with the non-linear
evolution and second order bias, the last two terms bring in the
effects of redshift distortions.

The new dynamical variables required for redshift-space are
the gradient of the tidal field
\be
    F_{ijk}(\r,t) = \nabla_i \nabla_j  \nabla_k \nabla^{-2} \delta_1(\r,t),
\ee
and the second-order tidal fields
\be
    \Pi^\delta_{ij}(\r,t) = (\nabla_i \nabla_j \nabla^{-2} -
    \frac{1}{3}\delta_{ij})\, \delta_1^2(\r,t)  ,
\ee
and
\be
    \Pi^E_{ij}(\r,t) = (\nabla_i \nabla_j \nabla^{-2}  -
    \frac{1}{3}\delta_{ij}) \,E^2(\r,t) .
\ee 
Note that in writing equations (\ref{final}) to (\ref{Del2}) we have
made the plane parallel approximation, evaluating all of the
redshift space contributions along a single cartesian axis. 

The remaining parameters in the second-order model are
(Bouchet et al. 1992, see also Catelan et al. 1995)
\ba
    		\kappa\equiv\frac{D_2}{D^2_1}
		&\approx& -3/7 \Omega^{-2/63} \hspace{0.5in}
		{\rm open\,\,\, universe} \nn
		&\approx&-3/7 \Omega_m^{-1/143}\hspace{0.45in}
		{\rm flat\,\,\, universe}
\ea
where $\delta_1(\x,t)=D_1(t)\varepsilon_1(\x)$, and
$D_1(t)\approx(1+z)^{-\Omega^{0.6}}$ is the linear growth function
for an open model (Peebles 1980). The redshift-space parameters
are (Peebles 1980, Lahav et al. 1991, Martel 1991)
\ba
    		f_1\equiv \frac{d \ln D_1 }{ d \ln a} 
		&\approx& \Omega_m^{3/5} \hspace{0.4in}
		{\rm open\,\,\, universe} \nn
		&\approx&\Omega_m^{5/9}\hspace{0.45in}
		{\rm flat\,\,\, universe}
\ea 
 and (Hivon et al. 1995)
\ba
		f_2\equiv \frac{d \ln D_2 }{ d \ln a} 
		&\approx& 2\Omega^{4/7}\hspace{0.35in}
		{\rm open\,\,\, universe} \nn
		&\approx& 2 \Omega_m^{6/11} \hspace{0.3in}
		{\rm flat\,\,\, universe}
\ea

Note that our expression for the redshift distorted and biased galaxy
distribution, equation (\ref{final}), does not agree with Hivon et al.
(1995), even taking into account their slightly different projection
into redshift space. Hivons et al.'s expression does not appear to
have the correct expectation value, i.e. their quantity $\langle
\delta^s \rangle \ne 0$.

\subsection{The distribution of initial fields}
Following the analysis of TW2000 we aim to find the
joint probability for the fields in equation (\ref{heavens}).
Defining the parameter vector 
\mbox{$\y
= (\delta_1,\etab,\g,\E,\Pi^{\delta}_{zz}, \Pi^{E}_{zz},F_{zzz})$},
the initial distribution function is given by
\be
P(\y)=\frac{1}{((2\pi)^{n}\mid\det{\C}\mid)^{1/2}} \exp{\left
(-\frac{1}{2} \y^t\C^{-1}\y \right)},
\label{mult}
\ee
where the covariance matrix is
\be \C = \lgl \y \y^t \rgl.
\ee
Taking the plane-parallel approximation for redshift-space distortions
represents a significant simplification
since any dependence of $\delta^{s}_g$ on the off-diagonal parts of
$\Pi^{\delta}_{ij}$, $\Pi^{E}_{ij}$ and $F_{ijk}$ is removed, reducing
the size of the required covariance matrix from $34\times 34$ to
$15\times 15$.

Equation (\ref{mult}) is an approximation. The variables $\Pi^\delta$
and $\Pi^E$ are not strictly Gaussian random fields as they are
generated from the square of linear fields. However, our definition of these
quantities as trace-free makes them uncorrelated with all of the other 
fields. If we then assume that their distribution is Gaussian, as we may
expect them to be as a result of the Central Limit Theorem, they
become statistically independent Gaussian fields. To
test our Gaussian approximation of the distribution of the $\Pi$-fields
we generated  random Gaussian $\delta$-fields and calculated
$\Pi^\delta_{zz}$. From this we estimated its distribution and found that
a Gaussian with variance $\sigma^2(\Pi^\delta_{zz})=8 \sigma_0^4/45$ 
(see equation \ref{cov}) was indeed a good approximation. Later on, in 
Section 2.5, we
shall see that these terms do not contribute to the skewness of the
final distribution to lowest order. Hence the effects of this approximation
will only be apparent in the higher moments and at higher order.

The non-zero elements of the covariance matrix are given by. 
\ba
\langle \delta^2 \rangle &=& \sigma^2_0, \hspace{1.3cm}
\lgl \eta_i \eta_j \rgl = \frac{1}{3}\sigma^2_1 \delta_{ij}, \nn
\lgl g_i g_j \rgl &=&  \frac{1}{3}\sigma^{2}_{-1} \delta_{ij},
\hspace{0.5cm}
\lgl \eta_i g_j \rgl = \frac{1}{3}\sigma^2_0 \delta_{ij}, \nn
\lgl E_{ij} E_{kl} \rgl &=& \frac{1}{15} \sigma^2_0
\left(\delta_{ik}\delta_{jl}+\delta_{il}\delta_{jk}-
\frac{2}{3}\delta_{ij}\delta_{kl}\right) \nn
\langle \Pi^{\delta}_{zz} \Pi^{\delta}_{zz} \rangle  & = & \frac{8}{45} \sigma^4_0 \hspace{0.6cm}
\langle \Pi^{E}_{zz} \Pi^{E}_{zz} \rangle = \frac{22}{135} \sigma^4_0 \nn
\langle g_i F_{jkl} \rangle &  = & \frac{1}{15}\sigma_{0}^2
    (\delta_{ij} \delta_{kl} + \delta_{ik}\delta_{jl} +
    \delta_{il}\delta_{jk}) \nn
\langle F_{zzz} \eta_z \rangle &  = & \frac{1}{5}\sigma_{0}^2  \hspace{0.8cm} 
\langle F_{zzz} F_{zzz} \rangle = \frac{1}{7}\sigma_{1}^2
\label{cov}
\ea
Evaluating the action of equation (\ref{mult}) we find
\ba
\y^t\C^{-1}\y & = & \frac{\delta^{2}}{\sigma_{0}^{2}} +
        \frac{3}{(1-\gamma_{\nu}^{2})}
        \left(\frac{\eta^{2}}{\sigma_{1}^{2}} +
        \frac{g^{2}}{\sigma_{-1}^{2}} -
        2\gamma_{\nu} \frac{\etab . \g}{\sigma_1 \sigma_{-1}} \right)
\nonumber \\
              & + &  15 \frac{E^2}{\sigma_{0}^{2}} +
    \frac{63}{4\sigma_{1}^2}\lll \eta_{z} - \frac{5}{3}F_{zzz} \rrr^2
\nonumber \\
    & + & \frac{45}{8\sigma_{0}^4}\lll \Pi^{\delta}_{zz}\rrr^2 +
    \frac{135}{22\sigma_{0}^4}\lll \Pi^{E}_{zz}\rrr^2,
\label{action}
\ea
with determinant
\be
\det{\C} = \frac{22}{7}\frac{2^5}{3^{15}5^9}\sigma_0^{32}\sigma_{1}^2 \frac
{(1-\gamma_{\nu}^2)^3}{\gamma_{\nu}^6}.
\ee
The correlation parameter is defined by
\be
\gamma_{\nu} \equiv \frac{\sigma_{0}^2}{\sigma_{1}\sigma_{-1}}.
\ee
which is the correlation coefficient of the velocity and gradient
density fields. The variances are defined as
\be
    \sigma_{n}^{2}(k) = D_1^2(t) \int_0^\infty \! \frac{dk}{2 \pi^2}
            \, k^{2+2n}  P(k) e^{-k^2 R_s^2},
\label{sigma0}
\ee
where $R_s$ is the smoothing scale.
For linear CDM power spectra $\gamma_\nu$ is found to lie in the range
$0.50< \gamma_\nu<0.65$ (TW2000) for a wide range of scales. TW2000
discuss the effect of varying $\gamma_\nu$ and find that it is very
small in the given range. We henceforth adopt the value $\gamma_{\nu} =
0.55$ for all subsequent plots. 

\subsection{Propagation of the density distribution function}

The multivariate Gaussian distribution for initial fields is
propagated to later times using the Chapman-Kolmogorov equation;
\be
P(\bog{\beta}) = \int d \bog{\alpha} W(\bog{\alpha}|\bog{\beta})
P(\bog{\alpha}).
\ee
where $W(\bog{\alpha}|\bog{\beta})$ is the transition probability from
the state $\alpha$ to $\beta$.  For a deterministic process, such as
the gravitational evolution of a system from the linear to non-linear
regimes, the transition probability is given by a Dirac delta function
\be W(\delta^{s}_{g}|\y) = \delta_{D}\left[
\delta^{s}_{g} - \delta^{s}_1 - \delta^{s}_2 - \Delta^{s}_1 -
\Delta^{s}_2 \right].
\ee
The advantage of using the Chapman-Kolmogorov approach is that is
generalizes the change of variables to allow for stochasticity in the
transformation between initial and final states. This will be further
discussed in \S 2.4.

The probability distribution function for $\delta^{s}_{g}$ can be
written as an expectation over the stochastic variables $\y$
\be 
P(\delta^{s}_{g}) = \left\langle \delta_{D}(\delta^{s}_{g} -
\delta^{s}_1 - \delta^{s}_2 - \Delta^{s}_1 - \Delta^{s}_2)
\right\rangle_y 
\ee
The Characteristic Function for $P(\delta^{s}_{g})$, defined by
\be
\fg(J) \equiv \int_{-\infty}^\infty
        d\delta^{s}_g P(\delta^{s}_g)\exp \left(-iJ\delta^{s}_g\right)
\ee
and with inverse
\be
    P(\delta_g^s)= \int_{-\infty}^\infty
        \! \frac{dJ}{2\pi} \, \fg(J)\exp \left(iJ\delta^{s}_g\right)
\label{invgf}
\ee
can be expressed as the expectation value
\be
\fg(J) = \left\langle \exp{-iJ(\delta^{s}_1 + \delta^{s}_2 + \Delta^{s}_1
+  \Delta^{s}_2)}\right\rangle_y.
\label{expect}
\ee

As in TW2000, equation (\ref{expect}) reduces to a set of Gaussian type integrals
which yield the probability generating function, $\fg(J)$, for a smoothed
galaxy density field in redshift space;
\ba
\fg(J) & = & \prod_{i=1}^5 \theta_i^{-1/2}
    \exp{\left[-\frac{1}{2}iJ\sigma_0^2 b_2\right]}
    \exp{\left[-\frac{J^2 \sigma_0^2 N^2}{2 \theta_1 }\right]}
    \nonumber \\
    & \times &
    \exp{\left[-\frac{2}{45} \frac{ J^2\sigma_0^2f_1^2}{\theta_3}
    \left(1-\frac{iJ\sigma_0^2 N H}{f \theta_1}\right)^2 \right]}
    \nonumber \\
    & \times & \exp{\left[-\frac{1}{2}J^2\sigma_0^8f_2^2\kappa^2
    \frac{1379}{1080}\right]}
\label{genfun}
\ea
where
\ba
\theta_1 & = & 1 + 2 i J \sigma_0^2 A  \nonumber \\
\theta_2 & = & \left(1 + \frac{4}{15} i J \sigma_0^2 B\right)^4 \nonumber \\
\theta_3 & = & 1 + \frac{4}{15} i J \sigma_0^2 B
 + \frac{8}{45} i J \sigma_0^2 f_1^2 + \frac{8}{45}\frac{J^2\sigma_0^4 H^2}
{2 \theta_1}   \nonumber \\
\theta_4 & = & \left(1 - \frac{2}{3} i J \sigma_0^2 b_1 +
\frac{1}{9}J^2 \sigma_0^4 b_1^2 \frac{(1-\gamma_{\nu}^2)}{\gamma_{\nu}^2}
\right)^2
 \nonumber \\
\theta_5 & = & 1 -
    i \frac{1}{5}J \sigma_0^2 \left( f_1^2 - \frac{10}{3}M \right) \nn
    &+&
     J^2 \sigma_0^4 \left[
     M \frac{(1-\gamma_{\nu}^2)}{9 \gamma_{\nu}^2}
    \left( M+\frac{18}{15}f_1^2 \right) +
    \frac{1}{21} \frac{f_1^4}{\gamma_\nu^2}
    \left( 1- \frac{21}{25} \gamma_\nu^2\right) \right] \nn
\ea
and
\ba
A & = & \frac{1}{3}(2 - \kappa)b_1 + +\frac{1}{9}f_1^2 + \frac{1}{3}b_1 f_1
- \frac{1}{9} f_2 \kappa + \frac{1}{2}b_2 \nonumber \\
B & = & \frac{1}{2}(1+\kappa)b_1 + \frac{1}{6} f_2 \kappa \nonumber \\
H & = & f_1 \left( b_1 +\frac{2}{3}f_1\right) \nonumber \\
M & = & b_1(1 +  f_1) \nonumber \\
N & = & b_1 + \frac{1}{3}f_1
\label{dibits}
\ea

The 1-pt PDF can then be found by numerical integration of equation
(\ref{invgf}). Equations (\ref{genfun}) to (\ref{dibits}) represent the
main analytical result of the this paper.

\subsection{Stochastic Bias Schemes}
\label{SEB}
Instead of the local deterministic prescription for galaxy biasing
described by equation (\ref{bias}) we may choose to implement a
stochastic Eularian biasing scheme. In stochastic biasing (Pen 1998,
Dekel \& Lahav 1999) the relation between the underlying density field
of dark matter and the galaxy density field is a random process with
joint distribution $P(\delta_{g},\delta)$.

If we assume that the biasing transition distribution, $P(\delta_{g},\delta)$,
is a bivariate Gaussian of the form
\ba
    P(\delta_{g},\delta) &=& \frac{1}{2 \pi [\sigma_b^2 +
    b_1^2 \sigma_0^2(1-r^2)]} \nn
    &\times&
    \exp - \frac{1}{2}
    \frac{\delta_g^2 +\delta^2 b_1^2 (1+\sigma_b^2/b_1^2 \sigma^2_0)
     - 2 r b_1 \delta_g \delta }{\sigma_b^2 +
    b_1^2 \sigma_0^2(1-r^2)},
\ea
where the covariances between the fields are given to first order by
\ba
    \lgl \delta_g^2 \rgl &=& b_1^2 \sigma_0^2 + \sigma^2_b, \\
    \lgl \delta_g \delta \rgl &=& b_1 r \sigma_0^2
\ea
where $\sigma_b$ is the variance of the random component determining galaxy
formation and $r$ is the correlation coefficient between the mass and
galaxy distribution, then the galaxy distribution function can be written
\be
    P(\delta_g) = \lgl P(\delta_g, f(\y)) \rgl_{y},
\ee
where $f(\y)= \delta_1 + \delta_2$ in real space. 
The characteristic function for galaxies is then
\be
    G_{\delta_g}(J) = \int dJ' G_\delta(J') G_{\delta_g, \delta}(J,J')
\label{stoc}
\ee
where
\be
    \ln G_{\delta_g, \delta}(J,J') =
    - \frac{1}{2} (J^2 \sigma_0^2 + J'^2 (b_1^2 \sigma_0^2 + \sigma_b^2)
    + 2 J J' b_1 r \sigma_0^2)
\ee
Equation (\ref{stoc}) can be evaluated numerically.

The distribution function of galaxies can also be evaluated by numerically
integrating
\be
    P(\delta_g) = \int_{-\infty}^\infty \frac{dJ}{2\pi}
    \fg_\delta(J) P(\delta_g,J)
\ee
where
\be
    P(\delta_g,J) = \int d \delta e^{iJ\delta} P(\delta_g,\delta)
\ee
is the partially transformed characteristic function. We shall explore
the effects of stochastic bias elsewhere, but note that it can be
naturally incorporated into our probabilistic scheme. For the
remainder of the paper we shall only discuss the deterministic scheme.

\subsection{Skewness of a biased density field in redshift space}

The skewness of the biased redshift-space galaxy distribution
can be found from the characteristic function, $\fg(J)$, by
differentiation. The moment parameters are defined as
\be
S_n = \frac{\langle \delta^n \rangle_c}{\sigma_0^{2n - 2}},
\ee
where the cumulants (or connected moments) of the field are given by
\be
    \lgl \delta^n \rgl_c = \frac{\pd^n}{\pd [i J^{n}] } \ln \fg(J=0).
\ee
In redshift space the skewness parameter becomes
\be
S^s_3 = \frac{\langle (\delta^s_g)^3 \rangle_c}{\sigma_s^{4}},
\ee
where $\sigma_s^2$ is the linear, redshift space variance of the
galaxy density field, $\delta^s_g$.

Taking the third derivative and setting $J=0$ we find the redshift-space 
skewness parameter;
\ba
S^s_{3}b_1^4 F^2 & = &
    4b_1^3 \left( 1 + \frac{2}{3} \beta_1 + \frac{3}{25} \beta_1^2 \right)
    +
    3 b_2 b_1^2 \left( 1 + \frac{\beta_1}{3} \right)^2 \nn
    &-& 2 \kappa b_1^3 (1 + \frac{1}{3} \beta_2 )
    \left( 1 + \frac{2}{3} \beta_1 + \frac{7}{75} \beta_1^2 \right) \nn
    &+&
    2 b_1^4 \beta_1 \left(1 + \frac{19}{15} \beta_1 + 
	\frac{3}{5} \beta_1^2 \right)
    + \frac{6}{25} b_1^4 \beta_1^4
\label{skew}
\ea
where $F(\beta_1)$ is the linear redshift space enhancement factor
for the redshifted variance given to lowest order by (Kaiser 1987)
\be
F =  1 + \frac{2}{3} \beta_1 + \frac{1}{5} \beta_1^2.
\label{boost}
\ee
and
\be
    	\beta_1 \equiv \frac{f_1}{b_1}, \hspace{0.3in} 
	\beta_2 \equiv \frac{f_2}{b_1}.
\ee
The variance in redshift space is just 
\be
\sigma_s^2 = b_1^2 \left(1 + \frac{2}{3} \beta_1 + \frac{1}{5}
\beta_1^2 \right)\sigma_0^2.
\label{sigs}
\ee
\begin{figure*}
\centering
\subfigure{\epsfig{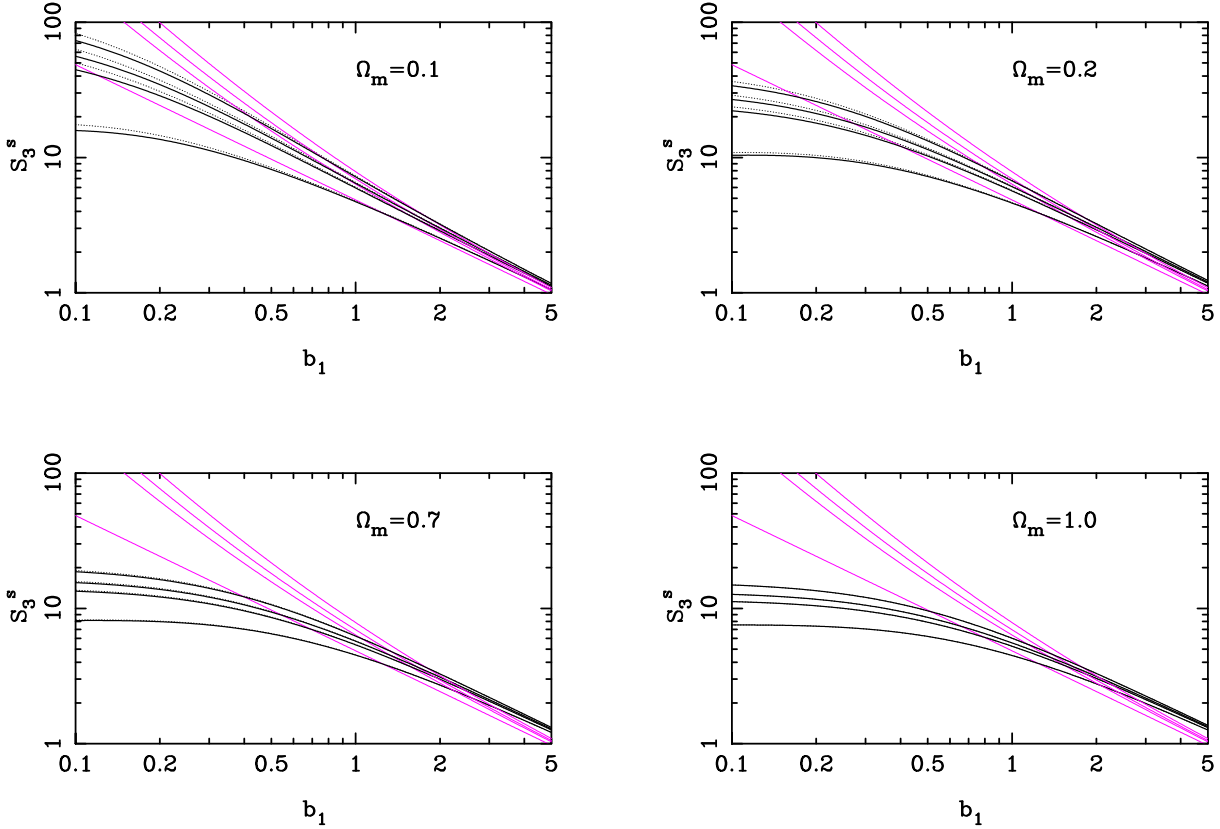}}
\caption{Log-log plot of the redshift-space galaxy skewness parameter, 
$S^s_3$, as a function of 
$\Omega_m$, $b_1$ and $b_2$. The panels show increasing values of 
$\Omega_m=0.1$, $0.2$, $0.7$, $1.0$. The x-axis is the linear bias 
parameter, $b_1$. Lines going up are the quadratic bias parameter 
$b_2=0.0$, $0.5$, $0.7$, $1.0$. Straight faint lines are the real space
skewness, $S_3$, and are independent of $\Omega_m$.
 Darker curves are for the full redshifted skewness 
for an open (solid) and flat (dotted) universe. }
\end{figure*}
Equation (\ref{skew}) for the galaxy skewness in redshift space is the
second major result of this paper. It is interesting to compare this
with the results of Hivon et al. (1995) who derived the density
skewness in redshift space. Hivon et al. derived their result by
transforming the redshifted bispectrum but did not find a closed form
solution and were left with a term containing an integral that was to
be determined numerically.  Our result for the skewness should
coincide with that of Hivon et al.  for the special case of $b_1=1$
and $b_2=0$, but does not.  The differences between our result and
that of Hivon et al. may result from the different approaches, or from
the difference in initial expressions for $\delta_g^s$. We have not
not been able to trace the origin of these differences. We have
verified equation (\ref{skew}) by a direct calculation of the
redshifted galaxy skewness, using equation (\ref{final}) along with
the correlators in equation (\ref{cov}), to evaluate the quantity $\langle
(\delta_g^s )^3\rangle$.

Equation (\ref{skew}) displays the hierarchical scaling associated
with galaxy clustering. It has already been shown that a local bias
will retain a hierarchical form if the underlying field displays
hierarchical behavior (Fry \& Gasta\~{n}aga 1993), and that
the lowest order of perturbation theory reproduces the hierarchical
structure. Our result would seem to indicate that this structure holds
for galaxies in redshift-space. This is an important result as
it strengthens the claim that if galaxies display a hierarchical
scaling then it is consistent that the underlying density field
was initially Gaussian distributed and bias is local. Our results
suggest that this can be extended as a test of galaxy distributions in
redshift-space.

In Figure 1 we plot the skewness parameter $S^s_3$ as a function of
linear bias, $b_1$, for different values of $\Omega_m$ and a quadratic
bias $b_2$. An immediate result is that in all the models the
redshift-space skewness parameter, $S^s_3$, is less than the
real-space skewness parameter, $S_3$, until $b_1>1.5$.  This may seem
surprising as one expects the effects of redshift space distortions
will tend to make the density field look more evolved along the line
of sight. This does happen, but the second-order skewness is dominated
by the first-order increase in the variance, so that the ratio $\lgl
(\delta^s)^3\rgl / \lgl (\delta^s)^2 \rgl^2$ is smaller than its real
space counterpart. This effect was previously observed by Hoyle et
al. (2000).

Beyond $b_1 \approx 1.5$ the redshift skewness parameter rises above
its real space value for all the models. This is because for large 
values of the linear bias parameter the second order skewness becomes
comparable in magnitude to the redshifted variance.

For $\beta_1\approx \beta_2 \approx 1$ and $b_1\approx 1$
the redshifted skewness can be approximated by
\be
    S_3^s = \frac{1539}{343} b_1^{-0.61} \beta_1^{-0.02} +
    \frac{75}{49} \frac{b_2}{b_1^2} \beta_1^{-9/14}
\ee
which is accurate to a few percent. This can be compared with the undistorted
skewness (Fry \& Gazta\~{n}aga 1993)
\be
    S_3 = \frac{34}{7} \frac{1}{b_1} + 3 \frac{b_2}{b_1^2}.
\ee
Without second-order bias, $S_3^s$ is only weakly dependent on $\beta_1$.
The main effect of redshift-space distortions in this case is to
change the dependence on $b_1$.  The second-order bias term, however,
has a much stronger dependence on $\beta_1$, and has a lower amplitude
than the undistorted term when \mbox{$\beta_1=1$}.

\begin{figure*}
\centering
\subfigure{\epsfig{figure=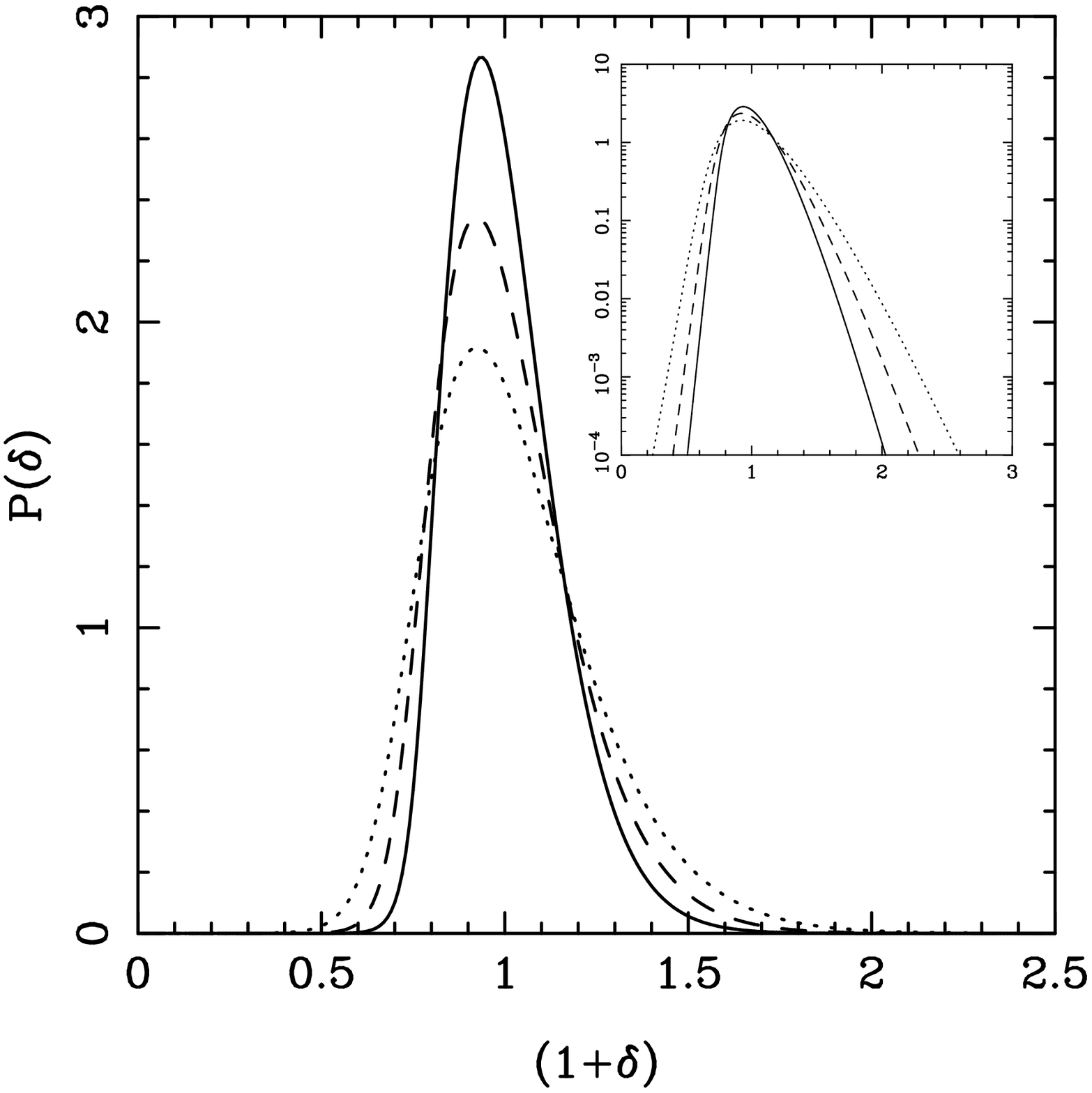,width=5.5cm,angle=0,clip=}}
\subfigure{\epsfig{figure=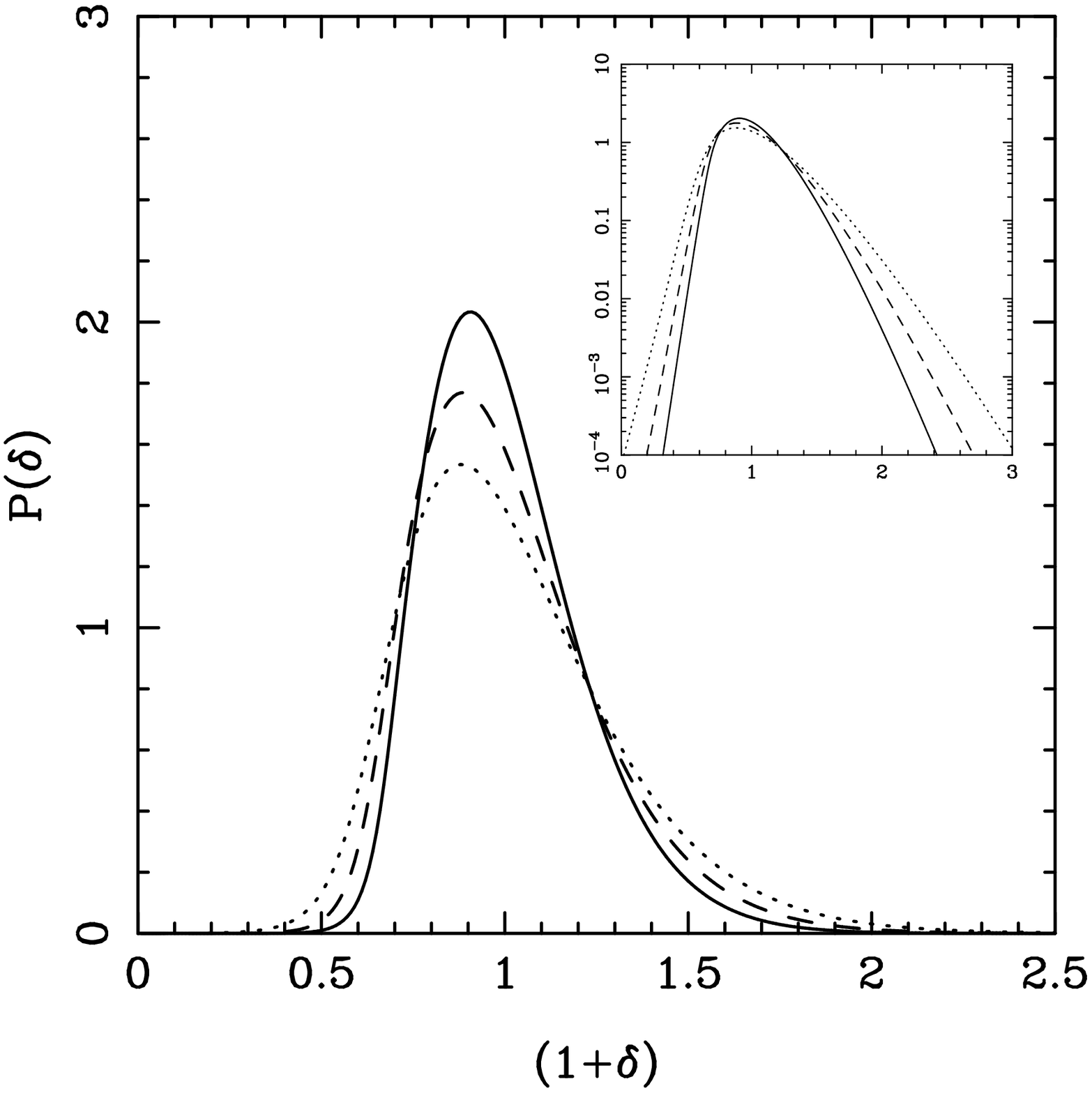 ,width=5.5cm,angle=0,clip=}}
\subfigure{\epsfig{figure=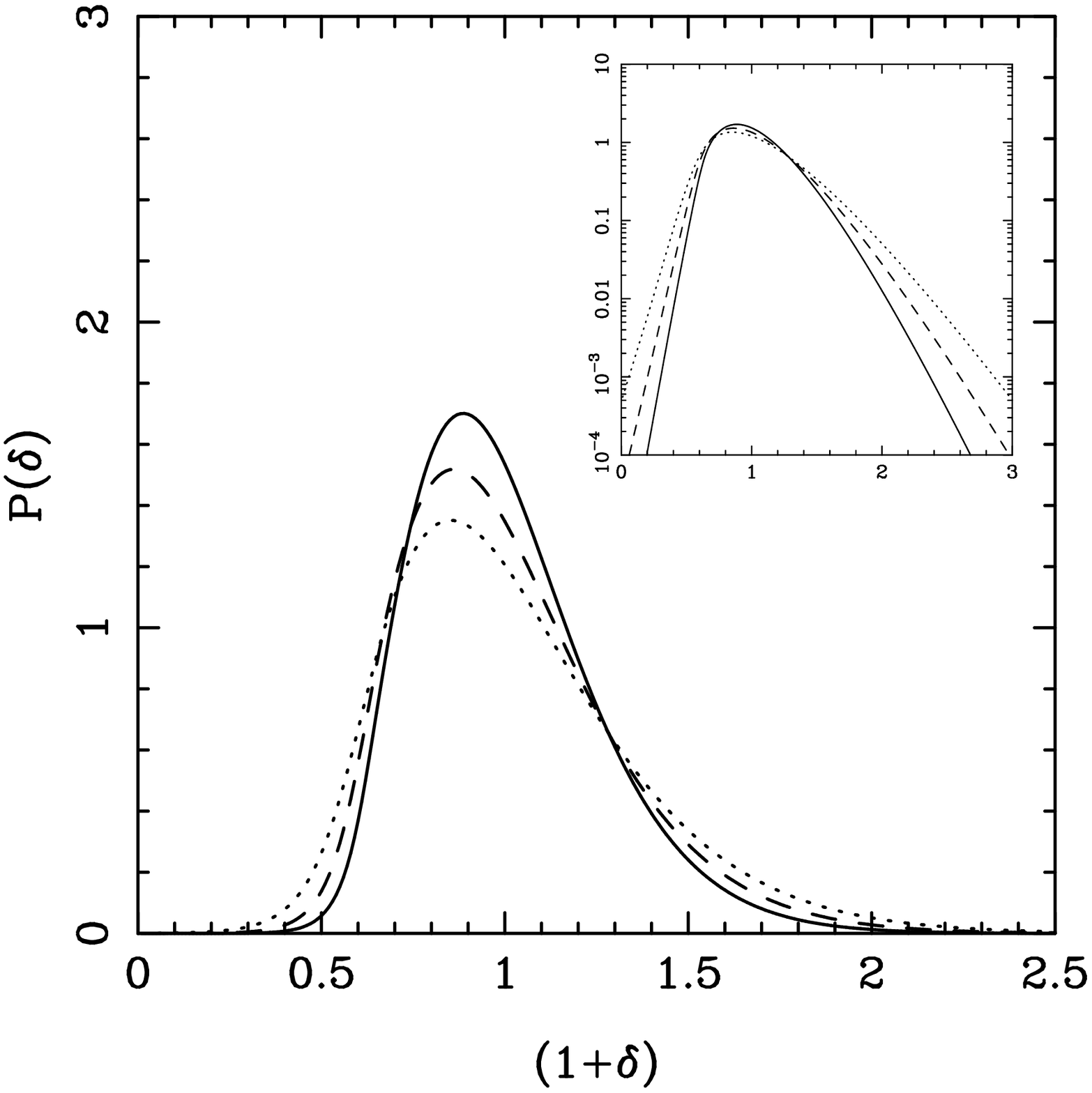,width=5.5cm,angle=0,clip=}}
\subfigure{\epsfig{figure=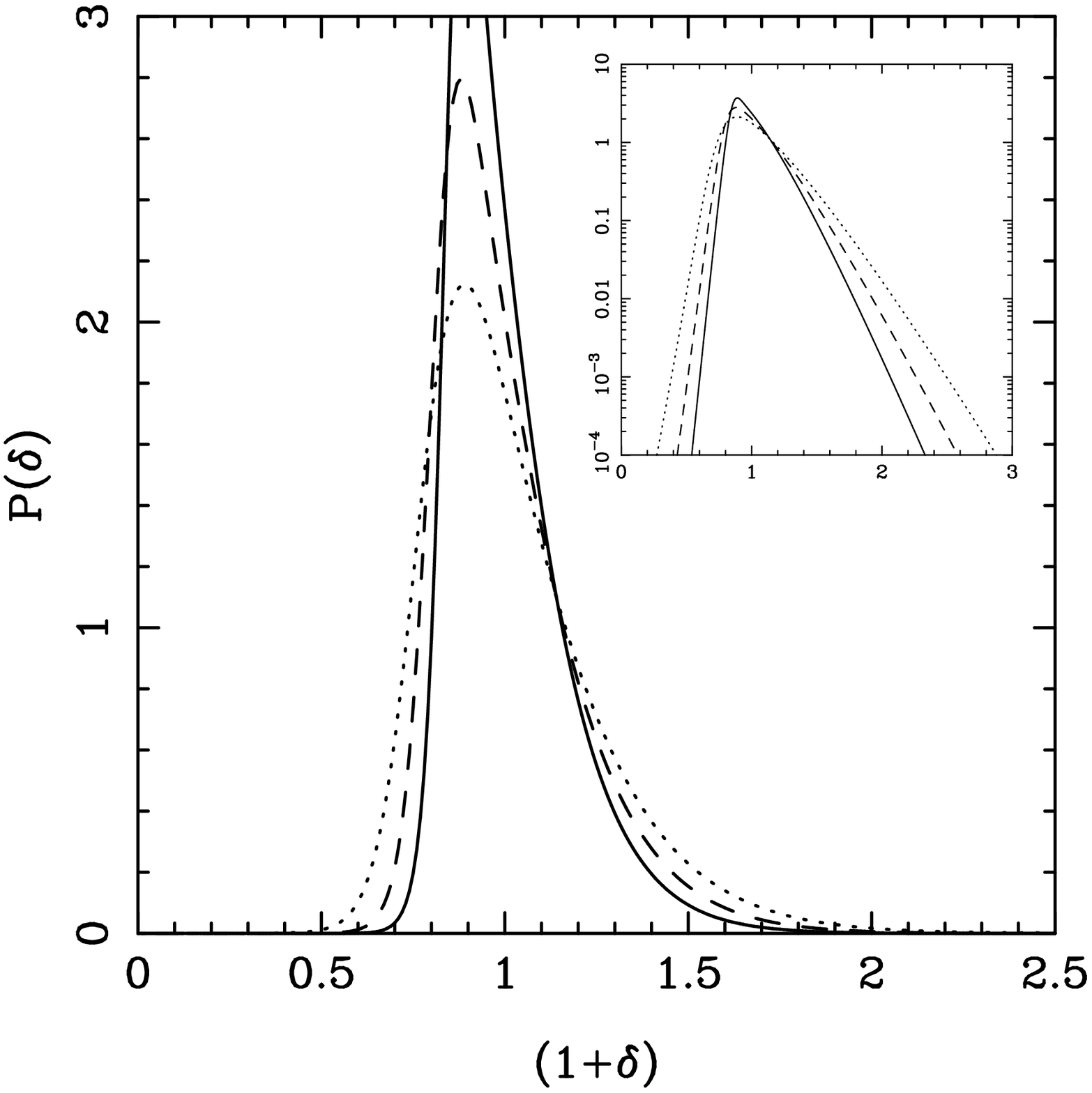,width=5.5cm,angle=0,clip=}}
\subfigure{\epsfig{figure=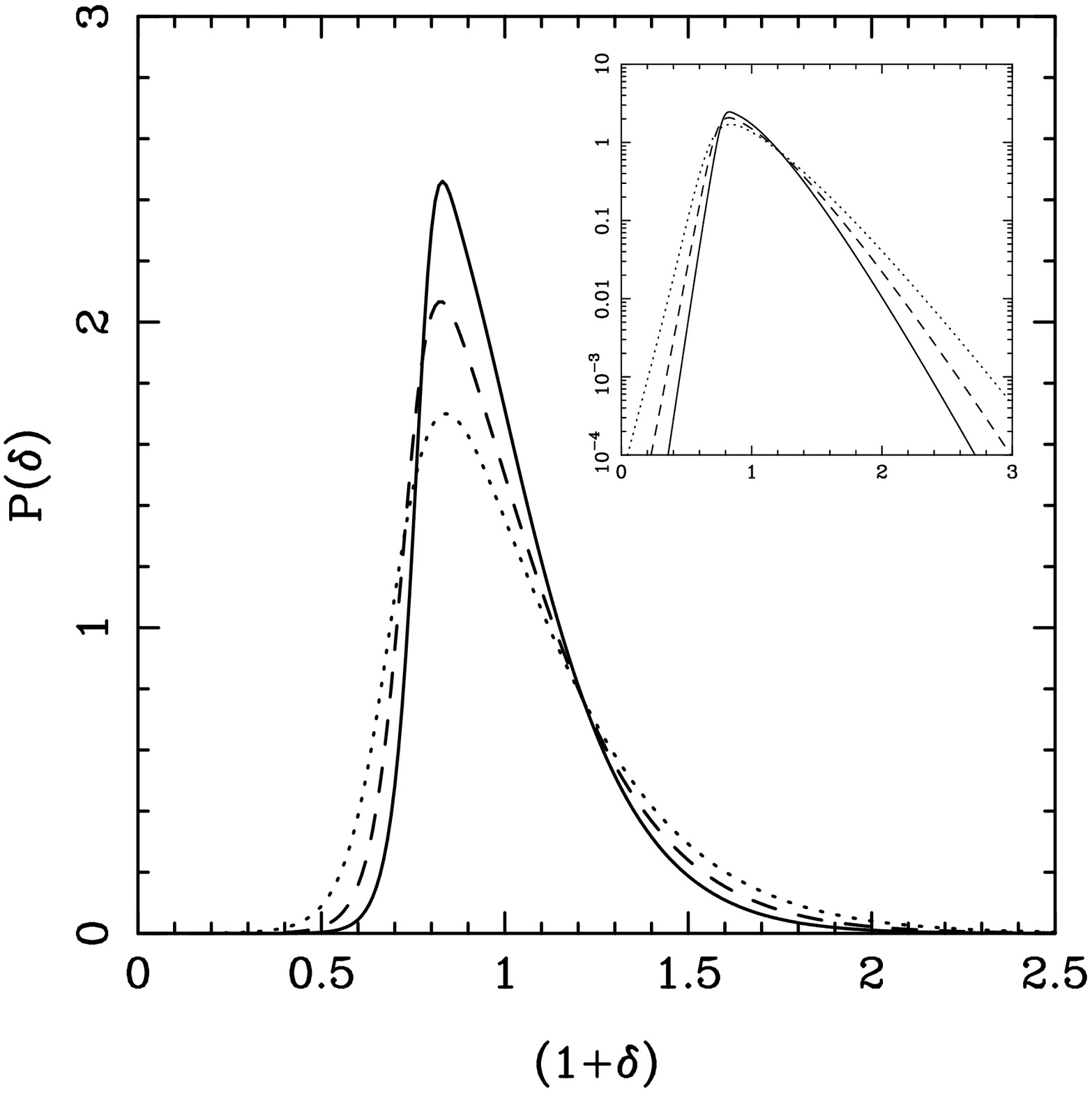,width=5.5cm,angle=0,clip=}}
\subfigure{\epsfig{figure=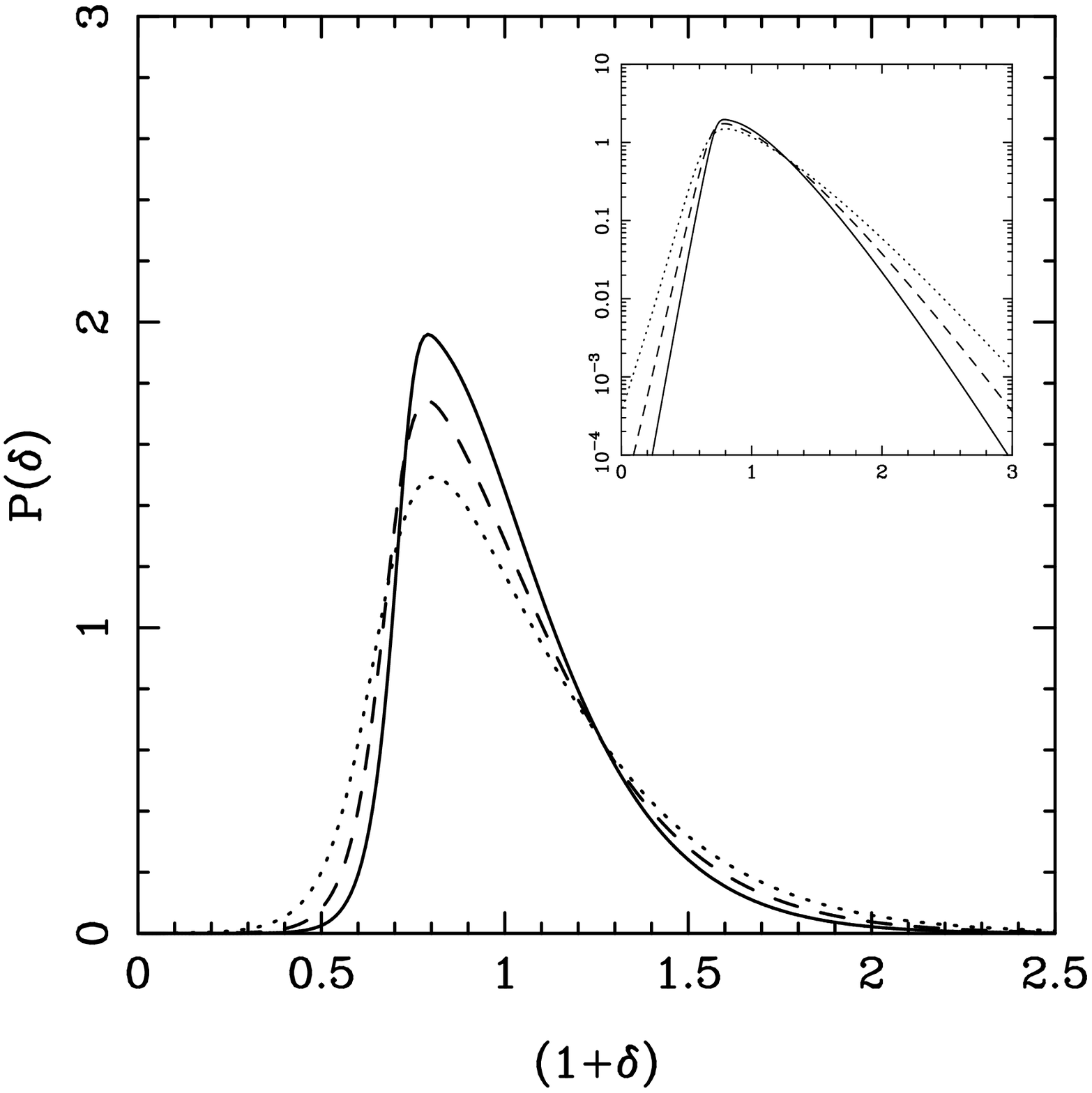,width=5.5cm,angle=0,clip=}}

\caption{The effect of linear and quadratic bias parameters on the
1-point PDF when the variance of the underlying mass density is
$\sigma_0 = 0.2$. From left to right the linear bias is $b_1 =$ 0.7,
1.0, 1.2. On the top row the second order bias parameter, $b_2$ is 0.0
and on the bottom 1.0. In all plots a solid line represents the PDF in
real space, a dashed line is for redshift space when $\Omega =$ 0.3, and
a dotted line for redshift space when $\Omega =$ 1.0. }
\end{figure*}

\section{The shape of the PDF}

The solution for the generating function, $\fg(J)$, given by
equations (\ref{genfun}) -- (\ref{dibits}) may be integrated numerically 
without too much
difficulty to yield the full distribution function $P(\delta)$. In
figure 2 we show the shape of the resulting PDF for a range of values
of the parameters $b_1, b_2$ and $\Omega_m$. 

To second order there are two important quantities working to distort
the shape of the PDF from a Gaussian with linear mass variance
$\sigma_0$. Firstly is the variance of the resulting field which acts
to broaden the distribution. Secondly is skewness, which
produces asymmetry in the PDF about its mean at $\delta = 0$. These
two quantities are effected to varying degrees by non-linear
evolution, by non-linear bias and by the distorting effect of galaxy
peculiar motions.

For the plots in figure 2 we chose a moderate value for the linear
variance of the underlying dark matter density field, $\sigma_0 =
0.20$. The correlation parameter, $\gamma_{\nu}$, was 0.55. In the top
panel we show the effect of a purely linear bias with values $b_1
=$0.7, 1.0 and 1.2 and with $b_2 = 0$. For the lower panel the
linear $b_1$'s match those above but in this case $b_2 = 1.0$ for every plot. 
Each plot shows three different PDFs: one (solid lines) in real
space, one in redshift space for the case $\Omega_0 = 1.0$ (dotted
lines) and a third in redshift space but with $\Omega_0 = 0.3$ (dashed
lines). We have inset the same plots on logarithmic axis in order to
emphasize the tails of the distribution.

In real space and with linear bias only, the shape of the PDF is
dominated by the $b_1^2$ boost to the variance. Where $b_1 > 1$ the
change to \sk is small and any effect on the shape of the PDF
caused by the skewness is masked by the enhanced variance
When $b_1 < 1$ there is a more substantial
change to \sk but in this case the variance becomes small enough that the
PDF looks relatively Gaussian regardless of what is happening to the
skewness.

To second order the non-linear bias parameter, $b_2$, does not alter
the variance of the underlying mass density field. The changes to the
shape of the PDF when $b_2 \ne 0$ then reflect only the corresponding
changes to \sk We observe that the real space PDFs appear to peak
sharply around $\delta = -1\sigma_s$ with an abrupt drop off of the low
density tail to the left of the peak. The high density tail is
extended beyond the corresponding case where $b_2 = 0$. With low $b_1$
the effect of $b_2$ is most pronounced, causing an increasingly abrupt
drop off in $P(\delta)$ on the low density side of the peak. This
pronounced effect on the PDF would appear to be an important signature
of quadratic bias, which cannot be reproduced by any combination of
other parameters.


\begin{figure*}
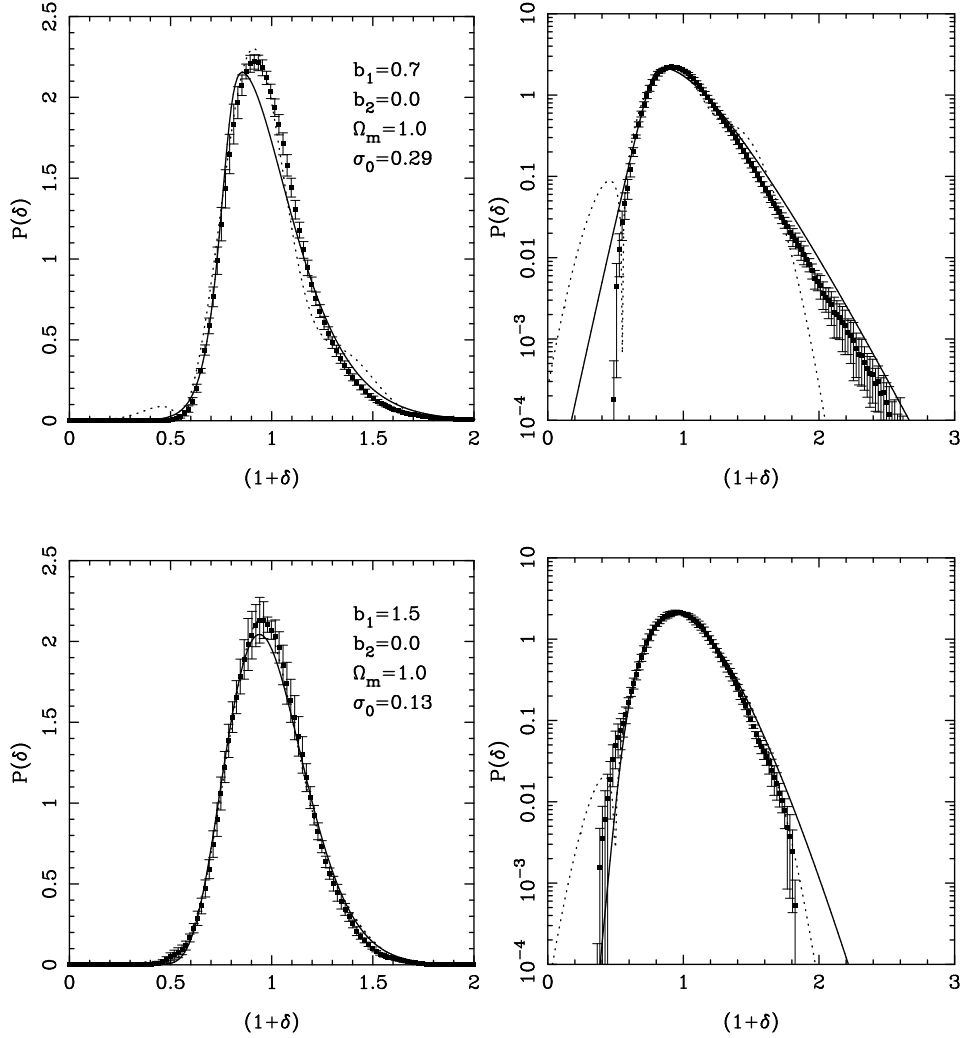

\centering
\subfigure{\epsfig{figure=fg1.ps,width=6.5cm,angle=270,clip=}}
\subfigure{\epsfig{figure=fg1log.ps,width=6.5cm,angle=270,clip=}}
\subfigure{\epsfig{figure=fg2.ps,width=6.5cm,angle=270,clip=}}
\subfigure{\epsfig{figure=fg2log.ps,width=6.5cm,angle=270,clip=}}
\caption{Comparison with N-body simulation in real space when
$\sigma_s = 0.20$ and with linear bias $b_1 = 0.7$ (top panel) and
$b_1 = 1.5$ (bottom panel). The solid line is the PDF for this work,
the dotted line shows the reconstruction of the PDF from the Edgeworth
expansion with the skewness parameter from \S 2. Filled squares
represent the simulation data. The linear mass variance was $\sigma_0
= 0.29$ and 0.13 for the top and bottom plots respectively. }
\end{figure*}

In redshift space the PDF is dominated by the enhancement of the variance
due to linear redshift space distortions. As discussed in \S 2.5 the 
second-order change in the skewness is less than the linear 
change in variance resulting in a lower skewness parameter, $S_3^s$, 
in redshift-space than in real space. The result
for the PDFs is a broader and less asymmetric appearance. This is most
clearly seen in the PDFs where $b_2 \ne 0$. For these plots, the sharp
peak in $P(\delta)$ around $\delta = -1\sigma_s$ is significantly
damped, and both the low and high density tails are much shallower
than their real space counterparts.

\section{Comparison of results with n-body simulations}
\subsection{The simulations}

In this section we illustrate how our theoretical PDF compares with
that measured from cosmological N-body simulations. For the comparison
we chose to use Hugh Couchman's Adaptive P$^3$M N-body code (Couchman
1991) to model the evolution of the dark matter. The simulation volume
was a cube of comoving side 200 $h^{-1}$Mpc with periodic boundary
conditions. We chose a CDM initial power spectrum (Bardeen at
al. 1986) that was normalized to match the present day abundance of
clusters. The variance on a scale of 8 $h^{-1}$Mpc when the particle
distribution was smoothed with top hat filters was $\sigma_8=0.5\
\Omega_{m}^{-0.5}$ in the final time-step of the simulation. The shape
parameter used was $\Gamma = 0.25$, representing the best fit of the
CDM power spectrum to galaxy clustering data. The simulation was
performed on a 128$^3$ Fourier mesh with $100^3$ particles.


\begin{figure*}
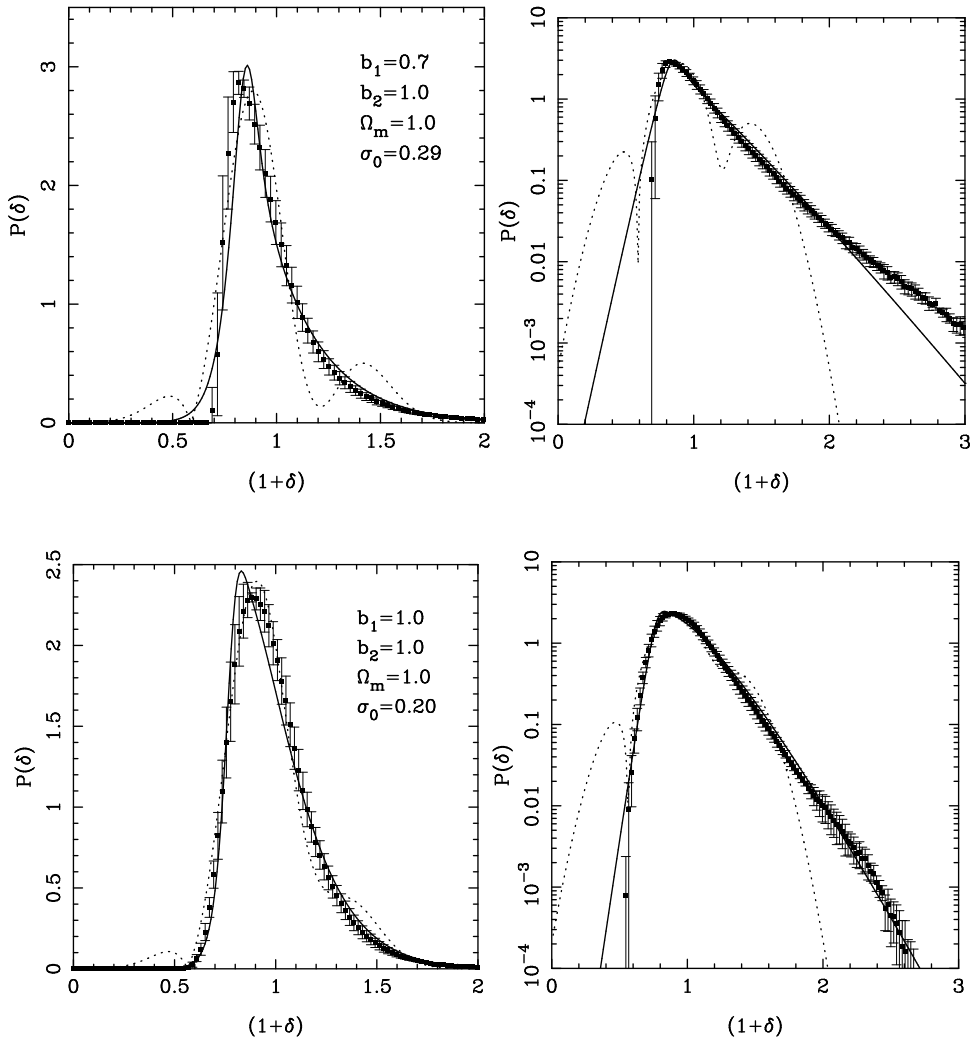

\centering
\subfigure{\epsfig{figure=fg3.ps,width=6.5cm,angle=270,clip=}}
\subfigure{\epsfig{figure=fg3log.ps,width=6.5cm,angle=270,clip=}}
\subfigure{\epsfig{figure=fg4.ps,width=6.5cm,angle=270,clip=}}
\subfigure{\epsfig{figure=fg4log.ps,width=6.5cm,angle=270,clip=}}
\caption{Comparison with N-body simulation in real space
when $\sigma_s = 0.20$ and
with linear bias and second order bias. In the top panel $b_1 = 0.7$
while in the lower panel $b_1 = 1.0$ In both cases the second order
bias parameter $b_2 = 1.0$. The solid line is the PDF for this work,
the dotted line shows the reconstruction of the PDF from the Edgeworth
expansion with the skewness from \S 2. Filled squares represent the
simulation data. The linear mass variance was $\sigma_0
= 0.29$ and 0.13 for the top and bottom plots respectively.}
\end{figure*}

We investigated the numerical PDFs in three sets of circumstances: 
\begin{itemize} 
\item in real space with bias, 
\item in redshift space without bias,
\item in redshift space with bias.
\end{itemize}  
The first two scenarios were trivial to construct
from the simulations.  
For the case of bias in real
space, we binned the data and estimated the overdensity from
$\delta=(n_p-\bar{n}_p)/\bar{n}_p$, where $n_p$ is the number of particles
in a cell and $\bar{n}_p$ is the mean number of particles per cell.
The biased distribution was then estimated by the transformation
\be
	\delta_g = b_1 \delta + \frac{b_2}{2}(\delta^2 - \sigma^2_0)
\ee
where $\sigma_0$ is the variance on the scale of the binning.

Redshift distortions, in the absence of bias,  were calculated using the
peculiar velocities of the simulation particles. The distortions were
made plane parallel in order to match our approximation from \S
2.1. The resulting particle distribution was smoothed with Gaussian
filters of radius $R_s$. The PDF was then evaluated from the relative
abundance of $\delta$ across the grid. 

Measuring the combined effect of redshift distortions and bias was
more difficult.  A simple combination of the above methods could not
be used, as we needed to identify biased galaxies before making the
transformation to redshift space. Our basic biasing method calculates
the biased density field on the grid so that information about
individual galaxies was lost. In order to identify individual biased
galaxies in the simulation we sampled a population of galaxies from
the simulation particles so that the number of galaxies in a cell,
$n_g$, was 
\be n_g = \bar{n}_g \left[1 + b_1\delta +
\frac{b_2}{2}(\delta^2 - \sigma_0^2) \right].  
\ee 
The resulting distribution could then be transformed to redshift-space
by using the galaxies velocity.
\begin{figure*}
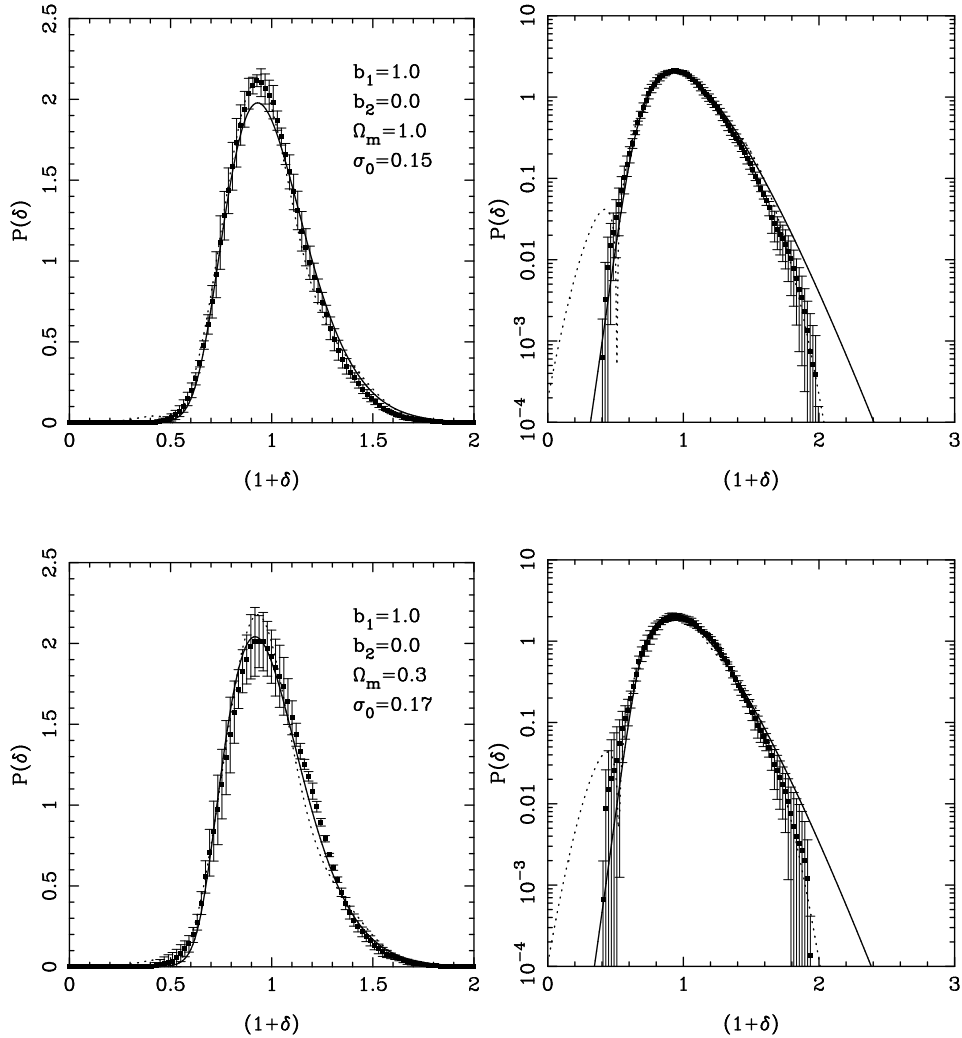

\centering
\subfigure{\epsfig{figure=fg5.ps,width=6.5cm,angle=270,clip=}}
\subfigure{\epsfig{figure=fg5log.ps ,width=6.5cm,angle=270,clip=}}
\subfigure{\epsfig{figure=fg6.ps,width=6.5cm,angle=270,clip=}}
\subfigure{\epsfig{figure=fg6log.ps,width=6.5cm,angle=270,clip=}}
\caption{Comparison with N-body simulations in redshift space for
$\Omega_0 = 1.0$ and $\Omega_0 = 0.3$ when $\sigma_s = 0.2$. The bias
parameters were in both cases $b_1 = 1.0$ and $b_2 = 0$, i.e no bias.
The solid line is the PDF for this work, the dotted line shows the
reconstruction of the PDF from the Edgeworth expansion with the
skewness parameter from \S 2. Filled squared represents the simulation
data. The linear mass variance was $\sigma_0
= 0.15$ and 0.17 for the top and bottom plots respectively. }
\end{figure*}

The size of the grid was dictated by the need for $n_p$ to be
sufficiently larger than $n_g$ so that underdense regions could be
enlarged by the bias. Given this constraint we chose to smooth the 
particle distribution on a coarse mesh of between 9 and 11 cells, 
with a cell size of $l=22\,h^{-1}{\rm Mpc}$ and $18\,h^{-1}{\rm Mpc}$,
respectively. As we calculate the linear variance using a Gaussian
filter (equation \ref{sigma0}) this corresponds to a smoothing
scale of $Rs=l/\sqrt{12}=6.5$ and $5\,h^{-1}{\rm Mpc}$.
 The total number of galaxies was
constrained by the requirement that $n_g$ was not larger than $n_p$ in
areas of high density, so we set the total number in the simulation 
volume $N_g=5\times10^{5}$.

The errors on the numerical PDFs for the biased and distorted only
simulations were taken from the standard deviation over 5 independent
realisations of the simulated volume.  Errors due to shot noise were
incorporated into the theoretical PDF (TW2000) based on the mean
density of particles in a cubical cell of side $l=\sqrt{12}R_s$. In
practice the shot noise contribution was small for simulations due to
the high particle density.

In the case of a combined biased and redshift-space distorted
PDFs the need to sample the galaxies on a coarser mesh meant that the
resulting PDF from an individual simulation had a larger scatter
from simulation to simulation for the same value of $\sigma_s^2$.
To avoid this we used a simulation with a smaller clustering variance.
 A relatively large number of particles could then to be found
in each cell. For the case $R_s \approx 5 h^{-1}$Mpc
there numbered on average 375 galaxies per cell sampled from around 750
simulation particles. When the numerical
PDFs were averaged over 7 independent realisations the results were
reasonably smooth, though the error bars (again the standard deviation
over the 7 realisations) do reflect the scatter.
\begin{figure*}
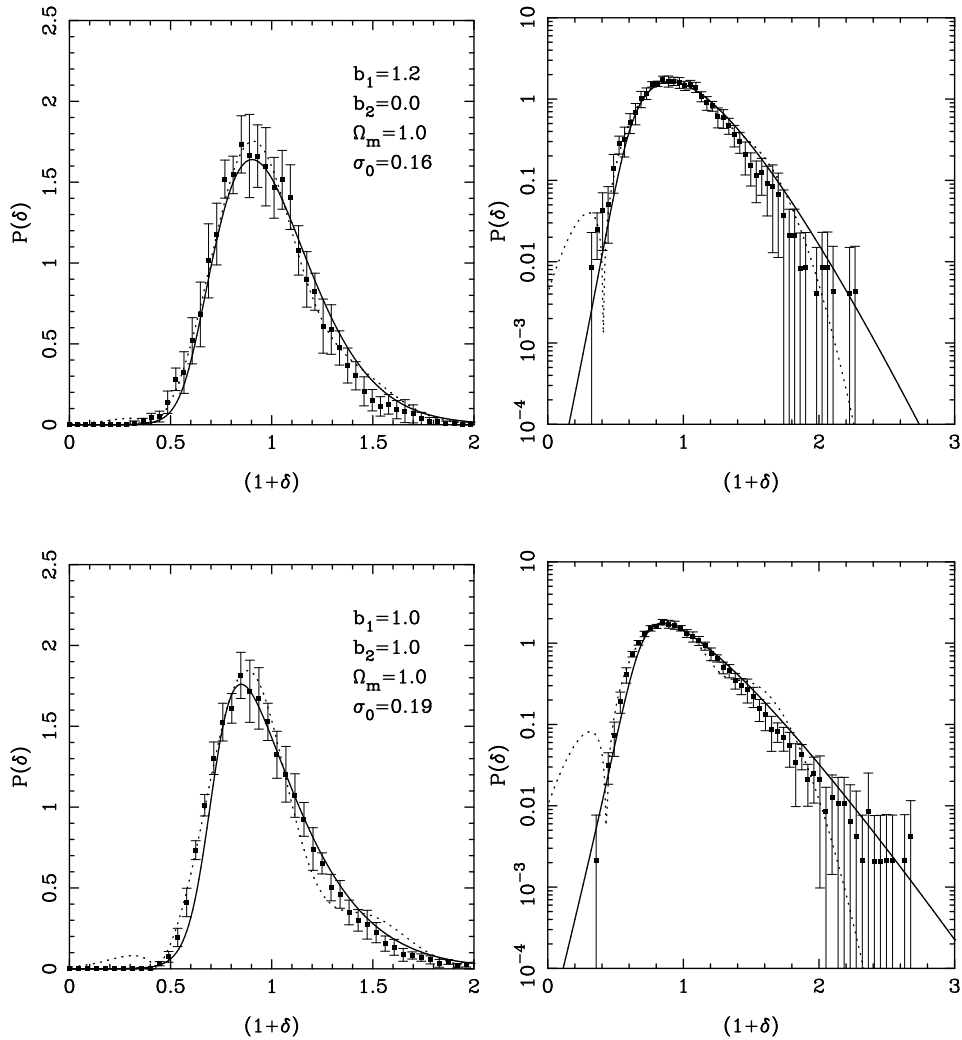

\centering
\subfigure{\epsfig{figure=fg7.ps,width=6.5cm,angle=270,clip=}}
\subfigure{\epsfig{figure=fg7log.ps,width=6.5cm,angle=270,clip=}}
\subfigure{\epsfig{figure=fg8.ps,width=6.5cm,angle=270,clip=}}
\subfigure{\epsfig{figure=fg8log.ps,width=6.5cm,angle=270,clip=}}
\caption{Comparison with N-body simulations showing the combined
effect of bias and redshift distortions for $\Omega_0 = 1.0$ and when
$\sigma_s = 0.25$. The bias parameters were $b_1 = 1.2$ and $b_2 = 0$
in the upper panel and $b_1 = 1.0$, $b_2 = 1.0$ in the lower.  The
solid line is the PDF for this work, the dotted line shows the
reconstruction of the PDF from the Edgeworth expansion with the
skewness parameter from \S 2. Filled squared represents the simulation
data. The linear mass variance was $\sigma_0 = 0.16$ and 0.19 for the
top and bottom plots respectively. }
\end{figure*}

\subsection{The Edgeworth Expansion}

In addition to the PDF calculated from this work we also examine
the PDF reconstructed from the Edgeworth expansion
(Juszkiewicz et al. 1995, Bernardeau \& Kofman 1995) using the
redshift-space skewness parameter derived in Section 2.5. The distribution 
function from the Edgeworth expansion is, to second order, 
\be 
P(\nu) = \frac{1}{\sqrt{2\pi}} \left[ 1 + \frac{1}{3!}H_3(\nu)
S^s_3 \right]\exp{\left[-\frac{1}{2}\nu^2\right]}
\ee 
where $\sigma_s^2$ is the redshift-space galaxy variance given by
equation (\ref{sigs}), $S^s_3$ is the skewness parameter from equation
(\ref{skew}), $H_n(\nu)$ is an $n^{th}$-order Hermite polynomial and
$\nu = \delta^s / \sigma_s$ is the scaled redshift-space density
field.

The main difference between the Edgeworth expansion and our own model
for the PDF is that the Edgeworth relies upon two levels of
approximation: firstly in the moments, which come from --- in our case
--- a second order perturbative calculation; and secondly in the
Edgeworth series itself, which reconstructs the non-linear PDF by
expanding about a Gaussian distribution. Our PDF comes directly from
the perturbation theory without any further constraints other than
those implicit in the theory itself. In this way our model is perhaps
more representative of the perturbation theory and its predictions and
limitations. Another dissimilarity between the models, and a problem
with trying to derive the PDF to second order, lies in the fact that
second order perturbation theory does not get the higher moments of
the distribution correct. In our calculation all of the higher moments
exist but have terms missing that would come in in a higher order
calculation. This may strongly effect the shape of our PDF,
particularly around the peak. Such problems are not an issue for the
Edgeworth and other similar approximations because in these models the
cumulants are either put into the series explicitly, or else are zero.

\subsection{Results}

In comparing the models with simulation data we constructed the PDF so
that the variance of the distorted, biased density field,
$\sigma_s^2$, was constant for all choices of the parameters $\Omega$,
$b_1$ and $b_2$. This was arranged by using an appropriate value for
$\sigma_0$ for each plot. For the case of bias in real space and of no
bias in redshift space we chose $\sigma_s = 0.2$. This relatively
moderate variance was necessary because for the cases where $b_1 < 1$
the underlying $\sigma_0$ had to start quite large. For the
combination of bias and redshift distortions we were limited to
starting with a slightly higher linear mass variance because the
simulation PDFs became dominated by sampling variance when the cell
size was large. For these plots $\sigma_s = 0.25$.

Figures 3 and 4 show the behavior of the PDF with various combinations
of the bias parameters and with no redshift space distortions.  The
density parameter for these simulations was $\Omega = 1.0$, though in
real space the quasi-linear evolution is not sensitive to this
quantity (Bouchet et al. 1992, Martel \& Freudling 1991).
 
The first set of plots (figure 3) show the effect of a linear bias
only, $b_1 = 0.7$ in the top panel and $b_1 = 1.5$ in the bottom. The
linear mass variance was $\sigma_0 = 0.29$ and $ 0.13$
respectively. The dominant effect on the shape is through the variance
which tends to broaden and lower the amplitude PDF for $b_1 > 1$ and
conversely makes the PDF more Gaussian when $b_1.  < 1$. The fit to
the data (points) of the PDF from this work is good in both cases,
though when $\sigma_0$ is relatively large our model misses the peak
of the distribution and appears to be slightly too strongly skewed. 
The Edgeworth approximation (dotted lines) also shows excellent
agreement with the simulations, particularly in the vicinity of the
peak. The plots show the absolute value of the Edgeworth PDF. 
The ``lobes'' in the low density tail of the
distribution represent negative probabilities in the Edgeworth PDF. 

In figure 4 we show the fit to simulations when the second order bias
term, $b_2$, is 1.0. The linear bias was set to $b_1 = 0.7$ (top) and
$b_1 = 1.0$ (bottom). The linear mass variance was $\sigma_0 = 0.29$
(top) and $\sigma_0 = 0.20$ (bottom). To the order we are interested
in $b_2$ only contributes to the PDF via $S_3$ and the higher
moments. 
The effect of the second order bias is to fill in the void
regions and enhance the peaks.
The result for the
PDF is a sharp drop-off of the low density tail and a dramatic amplification 
of the peak just beyond that point. 
The high density tail is also extended to account for
the increased regions of very high density. The fit to the simulations
for second order biasing is good so long as $\sigma_0$ is reasonably
small, though still in the quasi-linear regime. The breakdown in the
fit comes mainly in the low density tail which for the theoretical
curves drops away too slowly when the variance is high. The Edgeworth
PDF breaks down quite badly in the tails when $b_2 \ne 0$. This
suggests that the Edgeworth approximation becomes rather unstable when
the skewness parameter is large, even when the variance is low. The 
peak of the Edgeworth PDF is not so badly affected, however, and fits 
the data nicely so long as $\sigma_0$ is not too high.

Figure 5 shows the fit of the PDF to an unbiased field of galaxies in
redshift space. We ran two sets of simulations with $\Omega_0 =$ 1.0
and 0.3. The linear variance used for the theoretical PDFs was
$\sigma_0 = 0.15$ and $0.17$. The fit to both our model and the
Edgeworth PDF is very good, although the linear variance in each case
was quite small to ensure that $\sigma_s = 0.2$. We found that our
approximation began to break down when the redshift space variance was
$\sigma_s = 0.4$. Although the Edgeworth followed the peak and low
density part of the distribution well to this variance and higher, its
high density tail became unstable at a much lower $\sigma_s$.
In both cases the PDFs look more Gaussian than in real space because
of the reduction in $S^s_3$. The theoretical plots are almost
identical in each case, as expected since we have arranged for the
variance of the redshifted field to be identical in each case and
anyway the skewness in redshift space is only very weakly dependant
upon $\Omega_m$ when $b_1$ is close to unity. The numerical PDFs are
marginally different, although we suspect that this is a numerical
effect as suggested by the large error bars around the peak of the
distribution for $\Omega_m = 0.3$.

On much smaller scales we would
expect there to be a real difference between the two PDFs as 
fingers of god, due to the pairwise velocities in clusters, become important.
Our second-order analysis does not allow for these strongly
nonlinear effects, and we restrict our analysis to scales large
enough that these effects are not important. The agreement between
theory and the simulations suggests that this is true. The effect
we would expect to see would be a decrease in the variance and
skewness of the PDFs in redshift-space.

Our final plots (figure 6) show the combined effect of both bias and
redshift distortions. In the top row $b_1 = 1.2$ with $b_2 = 0.0$ and
on the bottom we have set $b_1 = 1.0$ and made $b_2 = 1.0$. For both
cases the density parameter was $\Omega_m = 1.0$. In each of the plots
$\sigma_s = 0.25$ --- due to problems with creating the bias/redshift
simulations we were unable to satisfactorily measure the PDFs for
$\sigma_s = 0.2$. The linear mass variance was therefore $\sigma_0 =
0.18$ in the top plot and $0.19$ in the bottom. The fit in both cases
is very good --- although the Edgeworth approximation becomes slightly
unstable when $b_2 = 1.0$ because of the relatively high
skewness. Although the redshift distortions do wash out some of the
effect of the bias parameters, there are dependencies which are
reproduced by the models to a reasonable degree of accuracy. This is
encouraging for the purposes of constraining cosmological models using
the PDF and data from galaxy redshift surveys.

In general there is a good similarity between all of the numerical
PDFs and those found from either our analysis based on the
Chapman--Kolmogorov equation or the Edgeworth approximation with the
redshift space skewness parameter $S^s_3$. The Edgeworth expansion
clearly breaks down sooner in the tails of the distribution, and for
large values of the skewness parameter.  Considering that both models
use second order perturbation theory and have the same skewness and
variance it is perhaps confusing that they behave so differently with
various combinations of the parameters. However we must not forget
that the Edgeworth approximation and our model rely on
very different approaches to obtain the PDF as discussed in \S 4.2. 

\begin{figure}
\centering
\subfigure{\epsfig{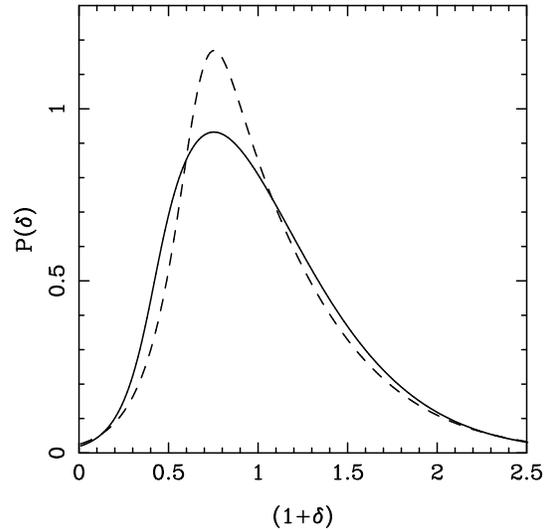}}
\subfigure{\epsfig{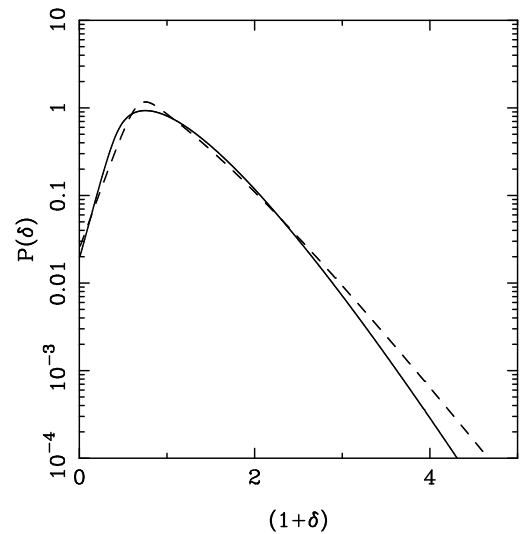}}
\caption{Theoretical PDFs for a biased galaxy field in redshift
space. The parameters are such that the combination $\beta = 0.5$. For
the solid curve $\Omega_0 = 1.0$ with $b_1 = 2.0$ and $b_2 = 0$ whereas
for the dotted curve $\Omega_0 = 0.3$ with $b_1 = 1.0$ and $b_2 =
0$. In both cases the filter was a Gaussian of radius 10
$h^-1$Mpc.}
\end{figure}

\subsection{Constant $\beta$ PDFs}

An interesting and potentially useful feature of modeling the PDF in
redshift space comes when one considers parameter combinations that
are degenerate to linear order. The combination 
\be 
\beta_1 = \frac{\Omega_m^{0.6}}{b_1}, 
\ee
for example, is a common quantity to measure in linear analysis of
galaxy redshift surveys (see e.g. Tadros et al. 1999 for a recent 
measurement from the PSCz redshift survey).

We constructed the PDFs of two fields sharing the same $\beta_1$ but
with different combinations of the cosmological density and linear
bias parameters, neglecting the effects of second order bias. We chose
$\beta_1 =0.5$ with $\Omega_m = 0.3$ and $b_1
= 1.0$ for one model and $\Omega_m = 1.0$ with $b_1 = 2.0$ for the
other. The linear mass variance for each was inferred from the
observed abundance of clusters from (Vianna \& Liddle 1996)
\be
\sigma(R_{th}) = 
\sigma_8 \left( \frac{R_{th}}{8 \, h^{-1} {\rm Mpc}}\right)^{-\gamma(R)}
\ee
where
\be
\gamma = (0.3\Gamma + 0.2)\left[2.92 + 
\log\left(\frac{R_{th}}{8\, h^{-1} {\rm Mpc}}\right)\right]
\ee
and with $\Gamma$ the CDM shape function. The abundance of clusters was parameterised by 
\be
\sigma_8 = 0.6 \Omega_0^{-C},
\ee
where $C \approx 0.49$ for the closed and $C \approx 0.43$ for the
open models.  The underlying linear mass
variance was found for each at constant top--hat filter radius,
$R_{th}$.  The resulting PDFs are shown in figure 6, the solid curve
being the high-$\Omega$ model and the dashed line the low-$\Omega$
model. Clearly there is a marked difference between the two PDFs, a
result of the different ways in which $b_1$ and $\Omega$ contribute to
the variance and the skewness. We investigate elsewhere whether this
difference can be used to constrain cosmological parameters.

\section{Summary}

In this paper we have derived an expression for the nonlinear 1-point
probability distribution of galaxies in redshift space. This function
is useful in the analysis of redshift surveys as it is a quantity that
can be directly observed. We have treated the nonlinearity of the
density field with second-order Eulerian perturbation theory, and
transformed from the mass-density distribution to a galaxy
distribution using a second-order local bias prescription. This is
then mapped to redshift space, again using second-order perturbation
theory. The transformation of the initial probability distribution to
the evolved distribution is carried out using the Chapman-Kolmogorov
equation. This allows us to derive an exact expression for the
second-order characteristic function for the galaxy distribution. We
show that the Chapman-Kolmogorov equation is more general than just a
transformation of variables, as we can use it to calculate the effects
of Stochastic Bias Schemes. We shall investigate this in more detail
elsewhere.

Taking the derivative of the galaxy characteristic function we derive
a new expression for the skewness parameter of galaxies in redshift
space, $S^s_3$.  Unlike other derivations, we find a closed-form
expression and include the effects of a quadratic bias term. We find
that in general for values of the linear bias parameter $b_1$ below
$1.5$, $S^s_3$ is smaller than its real space value, $S_3$. We also
find that, while the first-order bias terms are largely independent of
the linear distortion parameter $\beta_1$, or $\Omega_m$, for values
of $b_1$ around unity, as suggested by previous studies, the quadratic
bias terms introduce a strong dependency on cosmological parameters.
An analysis of the full 1-pt PDF in real and redshift space confirms
these findings. In addition we find that quadratic bias produces a
distinct sharp cut-of in the PDF for low-density regions. 

Comparing our PDF with those measured from N-body simulations we find
good agreement for all combinations of parameters. We also find that
the Edgeworth series, using our expression for the redshift-space
skewness parameter, fits the numerical PDFs rather well when the
series is truncated to second order and the redshift space skewness
and variance are used. The models, particularly the Edgeworth
approximation, fit the data most poorly when the linear mass variance
and the skewness are both high ($\sigma_0 > 0.4$). This occurs on small
scales or where there is a large degree of non-linear bias.

Our analysis leaves two problems unresolved. The first is the 
issue of fingers-of-god, which we do not model. This restricts
the scales on which we can model the PDF, but does not seem to be a
problem when we compare our results to simulations. We will test 
elsewhere if this is 
a problem when we come to use the PDF to extract cosmological parameters.
The second issue is that of smoothing the final density field.
We neglect this in our analysis, although it is included in the
analysis of other PDFs (see e.g. Bernardeau 1994). Again, when 
comparing with the results of simulations we do not see a significant 
effect. This may be a result of the type of power spectra we have 
tested our model against. The effect of smoothing is to transfer 
power across scales and for some power spectra this transfer 
will be minimal. However for the range of spectra we have investigated
we only require the variance to be correctly calibrated  against 
simulations and we find good agreement. An advantage of the PDF 
over other statistical measures is that it is straightforward to 
see if the fit is poor.

In future work the PDF we have derived will be compared with the PDF
of galaxies measured from the PSCz galaxy survey (Watts \& Taylor
in preparation). This will be useful both in testing
the gravitational instability hypothesis and in constraining
cosmological models, using data from the nonlinear regime of the
galaxy distribution. This step is vital for maximising the amount of
information available from galaxy redshift surveys.
Although we find that redshift distortions do
dampen some of the effects of bias, there are dependencies that may be
exploited. We have shown how using the PDF in conjunction with the
abundance of clusters may provide a way to distinguish between
cosmological models that are degenerate in linear analysis.

\section{Acknowledgements}
PIRW acknowledges the PPARC for a postgraduate studentship, ANT
acknowledges the PPARC for a postdoctoral fellowship.

\section{references}

\bib Alimi J.-M., Blanchard A., Schaeffer R., 1990, ApJ, 349, L5

\bib Bardeen J.M., Bond J.R., Kaiser N., Szalay A.S., 1986, ApJ, 304, 15

\bib Bernardeau F., 1992, ApJ, 392, 1

\bib Bernardeau F., 1994, AA 291, 697

\bib Bernardeau F., 1996, AA, 312, 11

\bib Bernardeau F., Kofman L., 1995, ApJ, 443, 479


\bib Bouchet F.R., Juszkiewicz R., Colombi S., Pellat R., 1992, ApJ, 394, L5

\bib Bouchet F.R., Strauss M.A., Davis M., Fisher K.B., Yahil A., Huchra J.P.,
	1993, ApJ, 417, 36

\bib Bouchet F.R., Colombi S., Hivon E., Juszkiewicz R., 1995, AA, 296, 575

\bib Bower R.G., Coles P., Frenk C.S., White S.D.M., 1993, ApJ 405, 403 

\bib Catelan P, Lucchin F., Matarrese S., Moscardini L, 1995, MNRAS, 276, 39


\bib Coles P., Jones B., 1991, MNRAS, 248, 1

\bib Coles P., 1993, MNRAS, 262, 1065

\bib Colombi S., Bernardeau F., Bouchet F.R., Hernquist L., 1997, 
	MNRAS, 287, 241

\bib Couchman H.M.P, 1991, ApJ, 368, L23

\bib Davis M., Peebles P.J., 1977, ApJ Suppl, 34, 425

\bib Dekel A., Lahav O., 1999, ApJ, 520, 24

\bib Fry J.N, Peebles P.J., 1978, ApJ, 221, 19

\bib Fry J.N, 1984, ApJ, 279, 499

\bib Fry J.N., 1985, ApJ, 289, 10

\bib Fry J., Gazta\~{n}aga E., 1993, ApJ, 413, 447

\bib Gazta\~{n}aga E., 1992, ApJ, 398, L17

\bib Gazta\~{n}aga E., 1994, MNRAS, 286, 913

\bib Gazta\~{n}aga E., Fosalba P., Elizalde E., 2000 (astro-ph/9906296)

\bib Groth E.J., Peebles P.J., 1977, ApJ , 217, 385

\bib Hamilton A.J.S., 1985, ApJ, 292, L35

\bib Hamilton A.J.S., 1998, in Hamilton D., ed., Ringberg Workshop on Large
Scale Structure 1996, The Evolving Universe. Kluwer Academic, Dordrecht
(astro-ph/9708102)


\bib Heavens A.F., Verde L., Matarrese S., 1998, MNRAS, 301, 797

\bib Hivon E., Bouchet F.R., Colombi S., Juszkiewicz R., 1995, AA, 298, 643

\bib Hoyle F., Szapudi I., Baugh C.M., 2000 (astro-ph/9911351)

\bib Hui L., Kofman L., Shandarin S.F., 2000 (astro-ph/9901104)

\bib Juszkiewicz R., 1981, MNRAS, 197, 931


\bib Juszkiewicz R., Bouchet F.R., Colombi S., 1993, ApJ, 412, L9

\bib Juszkeiwicz R., Weinberg D.H., Amsterdamski P., 
	Chodorowski M., Bouchet F., 1995, ApJ, 442, 39

\bib Kaiser N., 1987, MNRAS, 227, 1

\bib Kim R.S.J., Strauss M.A., 1998, ApJ, 493, 39

\bib Lahav O., Lilje P.B., Primack J.R., Rees M.J, 1991, MNRAS, 251, 128

\bib Martel H., 1991, ApJ, 377, 7

\bib Martel H., Freudling W., 1991, ApJ, 371, 1

\bib Peebles P.J., 1980, ``Large-Scale Structure in the Universe'',
   Princeton University Press, Princeton

\bib Pen U.-L., 1998, ApJ, 504, 601

\bib Press W.H., Schechter P., 1974, ApJ, 187, 425

\bib Szapudi I., Szalay A., Bosch\'{a}n P., 1992, ApJ, 390, 350

\bib Szapudi I., Meiksin A.,  Nichol R., 1996, ApJ, 473, 15

\bib Tadros H., Ballinger W.E., Taylor A.N., Heavens A.F., Efstathiou G., 
Saunders W., Frenk C.S., Keeble O., McMahon R., Maddox S.J., Oliver S., 
Rowan-Robinson M., Sutherland W.J., White S.D.M., 1999, MNRAS, 305, 527

\bib Taylor A.N, Watts P.I.R, 2000, MNRAS, in press (astro-ph/0001118)

\bib van Kampen, N.G., 
1992, ``Stochastic Processes in Physics and Chemistry'',
	North-Holland, Amsterdam

\bib Vianna P.T.P., Liddle A.R., 1996, MNRAS, 281, 323 

\bib Vishniac E.T., 1983, MNRAS, 203, 345

\end{document}